\documentclass[12pt, letterpaper]{article}
\usepackage[utf8]{inputenc}
\usepackage[utf8]{inputenc}
\usepackage[english]{babel}
\usepackage{amsmath}
\usepackage{amsfonts}
\usepackage{setspace}
\usepackage[ruled,vlined,linesnumbered,resetcount]{algorithm2e}
\usepackage[margin=1in]{geometry}
\usepackage{caption}
\usepackage{graphicx}
\usepackage{setspace}
\usepackage{natbib}
\usepackage{ntheorem}

\newtheorem{hyp}{Hypothesis}
\renewcommand\thehyp{\Roman{hyp}}

\makeatletter
\newcounter{subhyp} 
\let\savedc@hyp\c@hyp
\newenvironment{subhyp}
 {%
  \setcounter{subhyp}{0}%
  \stepcounter{hyp}%
  \edef\saved@hyp{\thehyp}
  \let\c@hyp\c@subhyp     
  \renewcommand{\thehyp}{\saved@hyp.\alph{hyp}}%
 }
 {}
\newcommand{\normhyp}{%
  \let\c@hyp\savedc@hyp 
  \renewcommand\thehyp{\arabic{hyp}}%
} 
\makeatother
\setcitestyle{authoryear}
\usepackage{authblk}

\title{Does Better Governance Guarantee Less Corruption? Evidence of Loss in Effectiveness of the Rule of Law}

\author[1,2]{Omar A. Guerrero}
\author[3]{Gonzalo Casta\~neda}
\affil[1]{The Alan Turing Institute, London}
\affil[2]{Department of Economics, University College London, London}
\affil[3]{Centro de Investigaci\'on y Docencia Econ\'omica (CIDE), Mexico City}

\date{}

\begin{document}

\maketitle

\begin{abstract}
Corruption is an endemic societal problem with profound implications in the development of nations. In combating this issue, cross-national evidence supporting the effectiveness of the rule of law seems at odds with poorly realized outcomes from reforms inspired in such literature. This paper provides an explanation for such contradiction. By taking a computational approach, we develop two methodological novelties into the empirical study of corruption: (1) generating large within-country variation by means of simulation (instead of cross-national data pooling), and (2) accounting for interactions between covariates through a spillover network. The latter (the network), seems responsible for a significant reduction in the effectiveness of the rule of law; especially among the least developed countries. We also find that effectiveness can be boosted by improving complementary policy issues that may lie beyond the governance agenda. Moreover, our simulations suggest that improvements to the rule of law are a necessary yet not sufficient condition to curve corruption.
\end{abstract}

\section{Introduction}\label{sec:intro}

In recent years, public governance has become one of the main topics in the international development agenda. However, in spite of significant efforts to improve governance through the rule of law,\footnote{The Oxford Dictionary defines rule of law as ``the restriction of the arbitrary exercise of power by subordinating it to well-defined and established laws''.} there seems to be a mismatch between the expectations from policy prescriptions and real-world outcomes.\footnote{These expectations were formed, to certain extent, by results from cross-national regression studies where the rule of law appeared, persistently, as one of the policy instruments with significant negative associations to aggregate corruption.} In this regard, the World Bank asserts --in its 2017 World Development Report: \emph{Governance and the Law}-- that legal improvements to the rule of law have rarely succeed in achieving drastic reductions of corruption.\footnote{Corruption is defined by the Oxford Dictionary as the ``dishonest or fraudulent conduct by those in power, typically involving bribery''. Another popular definition used by organizations such as Transparency International, the World Bank and the United Nations Development Program is ``the misuse of public office for private gain''.} \cite{baez-camargo_hidden_2017} offer a clue and one of the motivations for this study: that the ineffectiveness of reforms to the rule of law may originate from inconsistencies between the \emph{de jure} governance and the social norms that guide citizens and bureaucrats.

This paper studies, theoretically and empirically, a particular avenue for the ineffectiveness of the rule of law by means of a computational model. Its main contribution is new evidence of loss in effectiveness due to spillover effects to/from other policy issues. That is, while isolated improvements to the rule of law should, indeed, generate lower levels of corruption, such outcome is poorly realized because, in the real world, 1) the \emph{ceteris paribus} conditions for other policy issues do not hold and 2) co-movements in other topics introduce effects that may oppose the traditional conduits of anti-corruption policies (\emph{i.e.}, inverting the net benefit of misbehaving and curtailing the discretionary use of resources). For example, positive externalities to policy issue $i$ induce a new incentive structure in which the official in charge of $i$ has more opportunities to divert funds because its inflated performance (due to the externalitites) looks positive under imperfect supervision. 

Considering a non-\emph{ceteris paribus} setting with spillover effects allows us to move beyond the principal-agent framework and consider collective action mechanisms as a source of corruption. The latter has long been advocated for by several scholars in governance and public administration. Thus, we propose a new theoretical framework that can be illuminating for empirical studies of corruption and the discovery of abatement policies. In particular, we shed new light on the relationship between the rule of law and corruption through four types of hypotheses: (I) loss of effectiveness, (II) policy profile, (III) complementarity and (IV) priority. The loss-of-effectiveness hypotheses assess if improvements to the rule of law become ineffective under non-\emph{ceteris-paribus} conditions where spillovers are present. Policy profile hypotheses evaluate the presence of such impact when reforms to the rule of law are part of a larger policy package. The complementarity hypotheses test if policy issues with a negative association to corruption exhibit stronger effects when linked to the rule of law via spillovers. Finally, the priority hypotheses check if policy profiles that reduce corruption necessarily imply prioritizing the rule of law over other policy issues.

We produce country-level estimates from 115 countries during a sample period of 11 years and test these hypotheses. Our main findings consist of (I) evidence of a reduction in the effectiveness of the rule of law due to spillover effects; (II) a negative association between the rule of law and corruption not only in isolation, but as part of larger policy profiles; (III) that complementary policy issues with spillovers to/from the rule of law have significantly larger associations to corruption than those without; and (IV) that effective policy profiles do not translate into increasing the priority for the rule of law over other topics. The rest of the paper is structured in the following way. Section \ref{sec:study} provides a review of the literature. Section \ref{sec:data} introduces the data, the theoretical model and the empirical strategy. In section \ref{sec:results}, we present the empirical findings with regard to hypotheses family I. Section \ref{sec:policy} shows the results related to hypotheses families II, III and IV. Finally, we provide our conclusions in section \ref{sec:conclusions}.

\section{On the study of corruption and the rule of law}\label{sec:study}

\subsection{Principal-agent \emph{versus} collective action}

Broadly speaking, the empirical literature on the determinants of corruption tends to agree on the statistical significance of the rule of law (hereby called RoL), when tested in a cross-sectional setting. At the same time, however, there is substantial disappointment among international organizations with regard to the poor performance of institutional reforms inspired in such literature \citep[pp. 77-79]{worldbank_world_2017}. We argue that such discrepancy originates from a view that focuses exclusively in the principal-agent problem \citep{rose-ackerman_economics_1975,klitgaard_controlling_1988}. From such perspective, corruption arises from the presence of asymmetric information between the agents (\emph{i.e.}, public servants or elected officials) and a principal (\emph{i.e.}, government or voters) whose monitoring efforts are imperfect. Consequently, improvements to the RoL should reduce the agents' expected net benefits from diverting funds and curtail opportunities for the discretionary use of public resources.\footnote{This scenario occurs when the enforcement of the law increases the probability of catching offenders and when reforms to the RoL --and other governance mechanisms-- reduce the space for an unaccountable management of public funds. Broadly speaking, incentives for proper behavior are elicited through political competition, while rent-seeking opportunities are diminished by fostering economic competition \citep{ades_new_1997}.}

One of the problems with the principal-agent-only view is that systemic properties of corruption are considered irrelevant. This has been pointed out by \cite{persson_why_2013}, who argue for collective action as an account for corruption. In their view, the principal-agent model is ill-suited for explaining corruption because, in many developing countries, there are no principals willing to align the agents' interests with long-term societal welfare. Therefore, in countries without such principals, the expectation of corrupt behavior is widespread, reinforcing incentives to act in such manner. That is, when an individual believes that many others are corrupt, s/he does not have incentives to act differently. It is important to clarify that, in this scenario, dishonest behavior is not provoked by a lack of morality, but by a collective memory (or common knowledge) where high levels of corruption are socially tolerable. Because there is a generalized presumption that this is how society works, corruption becomes a collective-action problem.

In this paper, we intertwine the principal-agent and the collective-action perspectives. Our theory highlights two systemic features of corruption that are present in most nations: (1) an adaptive government that establishes policy priorities (resource allocations) across several policymaking offices; and (2) a spillover (positive externality) network among these policy issues (\emph{e.g.} health, education, infrastructure, public governance, etc.). From this perspective, we argue that corruption is the consequence of a political economy process through which the principal adapts to an uncertain environment, while agents learn to collectively establish social norms (corruption norms). Formally, we model such process as a game on a network. In this game, there is an information problem but, in contrast with the principal-agent view, this has to do with the uncertainty generated by the spillovers. Hence, the misalignment of incentives between government and bureaucrats gives place to a decentralized learning process.

\subsection{Econometric studies}

The econometric literature on the determinants of corruption is extensive and shows consensus with respect to the statistical significance of the RoL. The theory proposed in this paper aligns with this consensus in several ways, but differs in others. In particular, considering the interactions between different policy issues is not standard in these studies. Furthermore, due to data limitations, country-specific policy prescriptions are difficult to infer from traditional econometric frameworks. In this section, we review some seminal studies and elaborate on ways in which a computational approach could complement them.

Early studies on the determinants of corruption exploit the cross-national variation of different development indicators through pooled-regressions \citep{ades_new_1997,leite_does_1999,laporta_quality_1999,treisman_causes_2000,broadman_seeds_2001,dollar_are_2001,paldam_crosscountry_2002,fisman_decentralization_2002,herzfeld_corruption_2003,brunetti_free_2003,knack_trade_2003}. Overall, these studies have been consistent with the idea that governance instruments are effective tools that can be used in the fight against corruption. As the econometric literature has progressed, more sophisticated approaches have been deployed in order to overcome some of the limitations in these seminal works and provide a more fine-grained picture of the relevant policy tools.

In studies using Bayesian Model Averaging (BMA),\footnote{BMA is employed to deal with model uncertainty. A different approach, however, has been proposed by \cite{serra_empirical_2006} via Extreme Bound Analysis. Here, a predictor is considered robust when it remains statistically significant and preserves the same sign in all models that include such a variable. A less restrictive criterion for a predictor to be defined as robust is that the zero value is not included in the averaged 90\% confidence interval of the estimated coefficients \citep{seldadyo_determinants_2006}. The reader should be aware that this method is prone to multicollinearity problems due to potential interdependencies among the determinants. Hence, some auxiliary technique is required to cope with this issue.} \cite{gnimassoun_determinants_2016} and \cite{jetter_sorting_2018} find that some policy variables are robust predictors and, thus, they can be utilized by governments for abating corruption in relatively short periods. Some of these predictors include \emph{quality of education}, \emph{female participation in parliament}, \emph{willingness to delegate authority}, \emph{freedom of the press}, \emph{burden of regulation}, \emph{absence of political rights}, \emph{property rights} and  \emph{rule of law} (at least in one of the statistical analyses presented). It is important to emphasize that institutional covariates have a prominent role in this set of explanatory variables.

\cite{jetter_sorting_2018} apply a variant of the BMA to consider endogeneity in a large set of independent variables, instrumented through their one-decade lagged values. They find that, out of 32 potential determinants of corruption across 123 countries, 10 are robust. Furthermore, they identify five determinants with direct policy instruments: \emph{years of primary education}, \emph{trade freedom}, \emph{rule of law}, \emph{federal system}, \emph{absence of political rights}. Note that the last three are associated to the country's governance framework. Consistent with most cross-sectional studies, the level of economic development (GDP per capita) is also significant.\footnote{When causality is considered, \emph{years of primary education} and \emph{GDP per capita} become the two most relevant factors, while the RoL is still robust but less prominent. An alternative methodology that deals with reverse causality and heterskedasticity is three-stage least squares, as done by \cite{croix_democracy_2011}.}

Using quantile regression in order to deal with parameter heterogeneity, \cite{billger_existing_2009} identify that improvements in democracy have a negative effect on corruption only among the 50\% most-corrupt nations. On the other hand, increments in government size have negligible effects among the most corrupt countries. In an alternative strategy, \cite{gnimassoun_determinants_2016} and \cite{jetter_sorting_2018} split the sample by geographical region and development status, respectively. The latter authors, for example, find that the RoL is prominent among developing countries (\emph{i.e.}, non-members of the OECD), implying that the effectiveness of legal accountability diminishes once the quality of the RoL has reached certain level.\footnote{On one hand, several cross-country studies find a significant RoL coefficient \citep{ades_new_1997,leite_does_1999,broadman_seeds_2001,brunetti_free_2003,herzfeld_corruption_2003,ali_determinants_2002,park_determinants_2003,damania_persistence_2004,croix_democracy_2011,iwasaki_determinants_2012,mendonca_corruption_2012,elbahnasawy_determinants_2012}. On the other, others find significant coefficients among alternative governance indicators; some related to current policies (\emph{e.g.}, \emph{government effectiveness}, \emph{decentralization}, \emph{freedom of the press}, \emph{federal system}, \emph{women in parliament}, etc.) and others associated to the origins of the legal system. For extensive reviews on corruption and their economic, institutional and historical determinants \citep{jain_corruption_2001,lambsdorff_measuring_2016,pellegrini_causes_2011}; \cite[Ch.~5]{seldadyo_corruption_2008}; \cite{dimant_causes_2018}} In this sub-sample, only two of the 11 robust predictors relate to governance (RoL and \emph{absence of political rights}) while two more are associated to some policy instrument (\emph{foreign direct investment} and \emph{government size}).

In spite of these commendable efforts, there are still empirical challenges that need to be addressed; some related to the course-grained nature of development-indicator data, and others to methodological issues that are inherent to the econometric study of aggregate relationships. Regarding data, development indicators, generally, do not allow the exploitation of within-country variation (unless an extremely narrow set of covariates is used). While cross-national variation is, then, the dominant factor, its results have a limited policy interpretations since the estimated coefficients correspond to a hypothetical country with the average characteristics of the sample. Another data issue comes from the Rodrik critique \citep{rodrik_why_2012} which points out that policy indicators are not independent random variables, but conscious and strategic decisions made by governments in an attempt to obtain specific goals. Thus, the choice of development indicators as exogenous variables might not be appropriate.

The main methodological limitation comes from the Lucas critique, rejecting the assumption  that, under regression analysis, the estimated effects during the sample period will still be valid in an out-of-sample evaluation.\footnote{In the neoclassical view, it is commonly argued that only `deep' parameters (\emph{i.e.}, associated to technology or preferences) can be invariant. Hence, the associated prescription has to be estimated though micro-founded functional relationships. However, several authors have pointed out that such prescriptions are built on a flawed diagnostic on how complex societies operate (see \cite{colander_complexity_2014} for more on this criticism).} For example, given previous evidence on parameter heterogeneity across income groups, a country's estimates are likely to shift as its economy develops. Hence, in order to try to overcome some of these challenges, we propose a computational approach.

\subsection{Proposed empirical approach}


In this paper, we take a computational approach and argue that agent-computing can help overcoming problems of reverse causality, non-linearity, parameter homogeneity, the Lucas critique and policy endogeneity. To show this, we employ a computational model of the policymaking process, that allows producing country-specific estimates of the effectiveness of the RoL in reducing the diversion of public funds. 

Micro-founded computational models have the ability of producing generative causality \citep{epstein_chapter_2006}; something that we exploit to estimate the effectiveness of the RoL via controlled experiments. Generative causality means that the micro-level social mechanisms from our theory of corruption are formally specified in an algorithm, acting as the data-generating process. Through these experiments, we study the incidence that exogenous government decisions have on the aggregate level of corruption. In addition, this approach allows considering the endogenous variation of other policy issues that affect or are affected by the RoL, facilitating the estimation of a loss in effectiveness due to spillovers (something not doable under a \emph{ceteris paribus} assumption). Finally, since the algorithmic nature of the model allows specifications at the micro and macro levels, it can deal with the problem of parameter instability under counterfactuals \citep{castaneda_evaluating_2018}.\footnote{Recently, general equilibrium models have been developed to deal with interdependencies between endogenous (\emph{e.g.}, economic development and corruption) and exogenous variables (\emph{e.g.}, quality of governance). Here, inference comes from theorems (\emph{e.g.}, \cite{blackburn_incidence_2006,blackburn_public_2011}), regression estimates testing theoretical propositions (\emph{e.g.}, \cite{croix_democracy_2011,haque_corruption_2009,aidt_governance_2008}) or simulations (\emph{e.g.}, \cite{dzhumashev_corruption_2014,barreto_endogenous_2000}). Although, this approach is explicit about the causal channels between governance structure and corruption, it also has several drawbacks. First, the outcomes and mechanisms from equilibrium models are not properly validated with empirical evidence. Second, in the associated regressions, governance factors such as the RoL are assumed to be exogenous instead of endogenous variables; neglecting important processes such as learning and collective action. Third, many equilibrium models that produce solutions through simulations have too many parameters, often calibrated through questionable criteria (\emph{e.g.}, by adopting parameter values used by another author in a study of a different country).}

\section{Data and method}\label{sec:data}

\subsection{Data}
We build a dataset with 79 development indicators for 115 countries. Broadly speaking, the indicators can be categorized into 13 development pillars that roughly correspond to the pillars of the World Economic Forum.\footnote{We choose this classification over alternative ones (\emph{e.g.}, Sustainable Development Goals or World Development Indicators) because it contains more explicit governance instruments than the others.} We combine indicators from various sources such as the World Economic Forum's Global Competitiveness Report, the World Development Indicators, and the Worldwide Governance Indicators (WGI); the last two produced by the World Bank. Each indicator has been normalized in the range [0,1], and adjusted so that higher indices reflect better outcomes. 

The 13 development pillars covered by the 79 indicators are the following: \emph{governance of firms}, \emph{infrastructure}, \emph{macroeconomic environment},  \emph{health}, \emph{education}, \emph{goods market efficiency}, \emph{labor market efficiency}, \emph{financial market development}, \emph{technological readiness}, \emph{business sophistication}, \emph{R\&D innovation}, \emph{public governance} and \emph{cost of doing business}. From the 8 indicators belonging to the pillar of \emph{public governance}, two are especially relevant to our model: \emph{control of corruption} and \emph{rule of law}. Both of them belong to the WGI database. These indicators reflect the perception of citizens, entrepreneurs and experts in the public, private and NGO sectors. Although perception-based indices have well-known limitations, they are still one of the best metrics used in corruption studies. As defined by the WGI, \emph{control of corruption} ``captures perceptions of the extent to which public power is exercised for private gain''. We use this indicator as a proxy of the quality of the monitoring efforts by the central authority. On the other hand, the indicator of \emph{rule of law} captures ``perceptions of the extent to which agents have confidence in and abide by the rules of society''.\footnote{Note that the World Economic Forum provides an indicator for the diversion of public funds. We have excluded this variable from the sample as it corresponds directly to the definition of corruption used in the proposed model. Thus, this variable has been used to calibrate the model parameters as in done in \cite{castaneda_how_2018}.} This is our main independent variable.

In order to facilitate the presentation of our results (not for estimation purposes), we divide the sample into four clusters. These clusters are obtained through the Ward's clustering method, which minimizes the similarities between development indicators within clusters and maximizes the differences between members of different clusters. Roughly speaking, the four clusters match the  World Bank's income groups: (1) high, (2) middle-high, (3) middle-low and (4) low. However, in contrast with standard income classifications, clustering algorithms take into account the overall structure of countries' development indicators. The sampling period is 2006-2016, and the countries included in each cluster are specified in Appendix \ref{app:countries}.

\subsection{The model}

In order to overcome the limitations of regression-based studies and to combine the principal-agent and the collective-action views, we employ a computational model of the social mechanisms that act as the process that generates of corruption and development \citep{castaneda_how_2018}. Originally, such model was created to estimate government priorities from development-indicator data by simulating the resource-allocation process across multiple policy issues. Through an exogenous spillover network, it takes into account the potential incentive structure that, according to our theory, may arise from positive externalities. Therefore, it can also be used to examine the loss in effectiveness of the RoL derived form the proposed theoretical mechanisms.

The model consists of a political economy game where the central authority (the principal) sets policy priorities in terms of resource allocations, while public servants (the agents) are in charge of implementing such policies. The agents may use the allocated resources towards the policies that will transform their corresponding development indicators, or they may divert part of these resources for a personal gain. These decisions are shaped by the spillovers, the monitoring mechanism of the principal, the quality of the RoL and the profitability of corruption according to the social norm. Consequently, aggregate corruption is an endogenous variable.  In this section, we present an overview of the equations that are directly connected to RoL intervention.

In a typical simulation, the model begins with a vector $I$ of initial values for different development indicators of a given country. Then, the government tries to transform them by increasing their levels until reaching a target vector $T$. Every period, the central authority determines a vector of policy priorities $P$ (the allocation profile), such that $\sum_{i}^N P_{i,t} = B$ for every period $t$ and a budget constrain $B$. Each entry $P_{i,t}$ corresponds to the resources allocated to policy issue $i$. These resources are given to a bureaucrat with the mandate of transforming the development indicator corresponding to policy issue $i$. It is here, during the implementation phase, where corruption in the form of diverted public funds takes place. This leads to the first ingredient of the model, the public servant's benefit function

\begin{equation}
    F_{i,t} = (I_{i,t} + P_{i,t} - C_{i,t})(1 - \theta_{i,t}f_{R,t}).\label{eq:benefits}
\end{equation}

Equation \ref{eq:benefits} captures the `incentive effect' associated to an enhancement in the quality of the RoL. On the one hand, a bureaucrat receives political status from good indicator performance ($I_{i,t}$). On the other, s/he also benefits from extracting a private rent by diverting funds through a lower contribution $C_{i,t}$ towards the policy issue. However, her/his benefits are dampened if s/he is caught diverting funds. Thus, $\theta_{i,t}$ is a binary variable representing the random event of being caught. Here, randomness means that the monitoring mechanisms of the central authority are imperfect. However, the likelihood of successful outcomes is linked to the quality of such mechanisms and to the endogenous social norm of corruption, which we explain below. Finally, $f_{R,t} \in [0,1]$ captures the quality of the RoL. The intuition is that a higher $f_{R,t}$ translates into less impunity and larger punishments. 

As mentioned above, the probability of catching corruption depends on the quality of monitoring $f_{C,t} \in [0,1]$ and a social norm. More specifically, if official $i$ diverts a disproportionately large amount of resources, s/he will stand out and will likely be exposed by society, for example, through a media scandal. Therefore, the probability of $\theta_{i,t}=1$ is given by

\begin{equation}
         f_{C,t}\frac{(P_{i,t}-C_{i,t})}{\sum_{j=1}^N (P_{j,t}-C_{j,t})},\label{eq:catch}
\end{equation}

The terms $f_{R,t}$ and $f_{C,t}$ are endogenously determined by the evolution of the corresponding indicators of \emph{rule of law} (\emph{x = R}) and \emph{control of corruption} (\emph{x = c}) through the link function $I_{x,t}/(\exp{1-I_{x,t}})$. This marks an important departure from previous approaches: the quality of the RoL is endogenous. That is, exogenous government decisions to change the RoL may come from setting a new target $T_i$ for its quality, however, the evolution of the indicator is endogenous. Thus, changes to the RoL indicator are rather the consequence of modifications in the government's objectives and, then, generated by the model through specific channels. This allows estimating the effectiveness of the RoL while accounting for potential parameter shifts, interdependencies with other indicators and inefficiencies arising from the policymaking process.

Together, the agents' contributions and the spillover network drive the evolution of the indicators. We specify this process through

\begin{equation}
        I_{i,t} = I_{i,t-1} + \gamma(T_i-I_{i,t-1}) \sum_j C_{j,t}\mathbb{A}_{ji} 
    ,\label{eq:propagation}
\end{equation}
where $\mathbb{A}_{ji}$ is the adjacency matrix representing the spillover network as a weighted directed graph with ones in the diagonal. Parameter $\gamma$ represents the overall quality of public policies in a country takes the values estimated by \cite{castaneda_how_2018}.

As we show ahead, equation \ref{eq:benefits} is subject to different sources of uncertainty. For this reason, modeling the officials' decisions as a rational optimization problem is ill-suited. Instead, we opt for a simpler and more realistic behavioral model: \emph{directed learning}. The intuition of this approach is straightforward: if $F_{i,t} > F_{i,t-1}$ then the bureaucrat $i$ reinforces her/his previous action (diverting more or less funds), otherwise s/he changes the level of her/his contributions in the opposite direction to that of the previous period (see Appendix \ref{app:equations} for the corresponding equations).

The first and most obvious source of uncertainty in the bureaucrat's benefit function is the stochastic element of the principal's monitoring efforts. The second is the evolution of the associated indicator. This is so because an indicator's growth also depends on the spillovers coming from the contributions of other officials. The third source is the change of priority $P_i$ that the central authority assigns to policy issue $i$. Such change originates from the government's behavior, which we model as an adaptive process.

The government determines its allocation vector through 

\begin{equation}
    P_{i,t} = B\frac{q_{i,t}}{\sum_{i}q_{i,t}},\label{eq:allocation}
\end{equation}
where $q_{i,t}$ is the propensity to prioritize policy issue $i$ in period $t$, and is given by

\begin{equation}
    q_{i,t} = (T_i-I_{i,t})(K_i+1)(1-\theta_{i,t}f_{R,t}),\label{eq:propensity}
\end{equation}
where $K_i$ is the out-degree of policy issue $i$ (the number of non-zero entries in row $i$ of $\mathbb{A}$) and represents a proxy that the government uses to rank policy issues according to their socioeconomic importance (how central the issue is in the country's context).

Equation \ref{eq:propensity} shows how government actions generate changes in allocations and, hence, uncertainty to the officials. First, conditional on $K_i$, the government prioritizes the most laggard policy issues. Therefore, as specific topics progress, government allocations shift to other issues. Second, the term $1-\theta_{i,t}f_{R,t}$ means that the government readjusts its allocation whenever it discovers corruption. The magnitude of such reallocation depends on the quality of the RoL. Therefore, we can say that, through the `allocative effect' improvements to the RoL can diminish the government's propensity to allocate funds to highly corrupt officials.

Putting the pieces together, we can say that a RoL intervention starts at the level of one of the targets, representing the government's exogenous intentions (\emph{e.g.}, an increase in $T_i$, where $i$ corresponds to the RoL). Then, it trickles throughout the system via three channels: (1) an `incentive effect' that modifies the trade-off in the agent's benefit function; (2) an `allocative effect' that modifies the opportunities for wrongdoing; and (3) a `spillover effect' that boosts the agent's political status by improving its indicator. While the incentive and opportunity effects fit well into the principal-agent view of corruption, the spillover effect is a systemic property that coincides with the collective-action view. Arguably, it is possible to generate theoretical explanations where these three effects work in opposite directions. For this reason, it is important that evaluations of RoL-related policies account for the specific context faced by each country.

Overall, for a given country, the model takes as inputs the initial and final values of the development indicators. The former are the country's initial conditions and the latter its targets. For this study, we use the country-specific spillover networks estimated in \cite{castaneda_how_2018}. The model simulates the dynamics of the indicators until they reach their targets. Then, for a single simulation, we obtain the temporal evolution of the amount of diverted funds $P_i-C_i$. We elaborate on the specific metrics in the following sections.

\subsection{Simulated interventions}

Instead of exploiting the cross-national variation corresponding to the indicators of RoL and corruption, our strategy consists of using the model to endogenously generate different within-country levels of corruption by exogenously varying the target $T_R$ (not the indicator) of the RoL. There are several reasons why the targets are better exogenous variables than indicators. First, development goals (or targets) often come from various processes that are not necessarily related to policy implementation (at least not as strongly as indicators), for example, campaign promises, international agreements, political consensus, societal pressures or even discretionary decisions. In contrast, empirical development indicators originate from the policymaking process (which involves inefficiencies such as corruption), so they are likely conflated across topics and with the dependent variable. Second, when providing a policy prescription from econometric studies, it is assumed that a change in indicators equates to a similar change in policy priorities. This is unlikely to be the case since policy priorities are endogenous variables from the policymaking process. At the same time, other effects such as spillovers may be driving the indicators' dynamics. Third, increasing targets does not imply a proportional effect in corruption (a common assumption in linear models that use indicators). Thus, using them as the exogenous variables allows us to account for the potential non-linearities and bottlenecks coming from the data-generating mechanism.

We divide the empirical analysis in two groups. The first (presented in section \ref{sec:results} and containing hypotheses from family I) focuses on the exogenous modification of the target of the RoL ($T_R$). That is, we study its effect in isolation from all other targets. Note, however, that this procedure does not imply a \emph{ceteris paribus} assumption for the indicators and allocations of all other issues.\footnote{For a given country, a retrospective simulation consists of running the model from its initial conditions until the indicators reach the target vector $T$. Here, we assume that the targets are the empirical final values of the indicators. A counterfactual simulation, on the other hand, can be performed by modifying specific targets such as $T_R$. Then, we run a simulation with a $T_R$ larger than the one from the retrospective estimation. Given that the only varying factor between both types of simulations is $T_R$, we can assert that the resulting difference in corruption is caused by a change in the government's objective with respect to the quality of RoL.} In the second group (presented in section \ref{sec:policy} and containing hypotheses from families II, III and IV), we investigate corruption when varying the entire target vector. This has an intuitive interpretation in terms of countries adopting comprehensive policy profiles such as the ones prescribed by multilateral organizations. Hence, we explore potential complementarities between the RoL and other policy issues.

\subsection{Aggregate corruption}

At this point, it is important to clarify that a simulation period $t$ does not correspond to time. Instead, it represents the occurrence of events such as the reallocation of resources or the punishment of corruption. For this reason, inter-period metrics provide information about the frequency with which certain events occur during the time span of the dataset. Thus, our measure of aggregate corruption consists of the sum of all events of diversion of public funds throughout the simulation, averaged across all policy issues. More specifically, we quantify corruption as

\begin{equation}
    D = \frac{1}{B}\sum_i \sum_t (P_{i,t}-C_{i,t}).\label{eq:totCorr}
\end{equation} 

This measure of corruption has two components: (1) the aggregate level of diverted public funds per event and (2) the frequency with which these events take place. The first component reflects the nominal dimension of this form of inefficiency, while the second captures its commonality through time. Due to the dual nature of this metric, enhancements to the RoL have static and dynamic effects. The former appears, for instance, when the incentive effect encourages public servants to divert less. The latter takes place when the allocative effect shrinks the size of the pie among corrupt officials but, at the same time, may increase the number of periods needed to reach the targets because there is less budget to transform the affected indicator. Due to this static-dynamic dichotomy, one would not necessarily expect less corruption from improvements to the RoL, at least not in a linear fashion.

We can summarize the estimation of aggregate corruption as follows:

\begin{enumerate}
    \item Given initial conditions and targets $T$, run the model until $I_{i,t} \approx I_{i,t-1}$ for every $i$.
    \item Compute $D$ through equation \ref{eq:totCorr}.
    \item Repeat the 1 and 2 to collect a Monte Carlo sample (1000 simulations) of $D$.
    \item Compute the average $\bar{D}$ from the Monte Carlo sample.
\end{enumerate}

We obtain $S$ estimates $\{ \bar{D}_1, \dots , \bar{D}_S\}$, corresponding to the expected aggregate corruption for different sets of Monte Carlo simulations. For each set of simulations, we determine a different level for the target $T_R$ corresponding to the RoL. More specifically, we determine an increasing sequence of uniformly spaced targets $\{ T_{R_1}, \dots , T_{R_S} \}$ such that $T_{R_1}$ corresponds to the empirical one and $T_{R_S}=1$ which is the maximum value in the sample. Therefore, the estimates $\{ \bar{D}_1, \dots  , \bar{D}_S\}$ reflect the level of corruption generated by different degrees of improvement of the RoL with respect to the retrospective estimation.

An outcome that is consistent with the econometric literature should produce a negative relationship between $T_R$ and $\bar{D}$. However, should a non-linear relationship emerge, it would be the result of the micro-level data-generating process rather than the aggregate association between two variables: an endogenous variable (corruption) and a poorly conceived exogenous variable (RoL). Furthermore, the various effects previously explained can also give raise to a null relationship or even a positive one (but we would expect few). In addition, these estimates do not require pooling cross-national data, so any inference for policy purposes is based on the country's own experience.

\subsection{Loss of effectiveness}

In colloquial terms, effectiveness is understood as the reduction of corruption caused by improvements to the quality of the RoL. In econometric analyses, this is the interpretation of the coefficient corresponding to the RoL. More generally, let $f(T_R)$ be a function describing the relationship between $T_R$ and $\bar{D}$. Then, we can define the effectiveness of the RoL as $\frac{\Delta f}{\Delta T_R}$.

Let $g(T_R)$ represent another relationship between $T_R$ and $\bar{D}$; this time derived from a counterfactual analysis: omitting spillover effects. Then, the differences between $\frac{\Delta f}{\Delta T_R}$ and $\frac{\Delta g}{\Delta T_R}$ can give us an idea of the change in effectiveness of the RoL due to differences between the estimated and counterfactual worlds. These differences, however, might vary depending on the level of $T_R$ (especially in the presence non-linearities). Thus, we propose the following metric for the loss of effectiveness: 

\begin{equation}
L = 1-\frac{\int_{T_R} g(T_{R})dT_{R} - [g(T_{R_1})-f(T_{R_1})]} {\int_{T_R} f(T_{R})dT_{R}},\label{eq:index}
\end{equation} 
where subtracting $g(T_{R_1})-f(T_{R_1})$ allows controlling for nominal discrepancies in the levels of corruption (\emph{e.g.}, $f(T_R) < g(T_R)$ by virtue of the accelerated convergence to targets caused by the spillovers between indicators).\footnote{Confidence intervals can be obtained by repeating the entire procedure to compute $\bar{D}$ and then $L$. However, at the time we write this paper, this approach may be computationally prohibitive for large samples of countries and indicators. Therefore, we construct bootstrap confidence intervals by resampling the Monte Carlo ensemble $\{ D_1, \dots, D_n\}_i$ corresponding to each level of $T_{R_i}$ and computing $\bar{D}$. After computing the re-sampled $\bar{D}$ for all target levels in both specifications (with and without spillovers), we compute $L$. Repeating this procedure generates the bootstrap sample $L$, which we then use to obtain the index intervals.}

\begin{figure}[ht]
    \centering
    \caption{Example of loss of effectiveness due to spillover effects}
    \includegraphics[scale=.75]{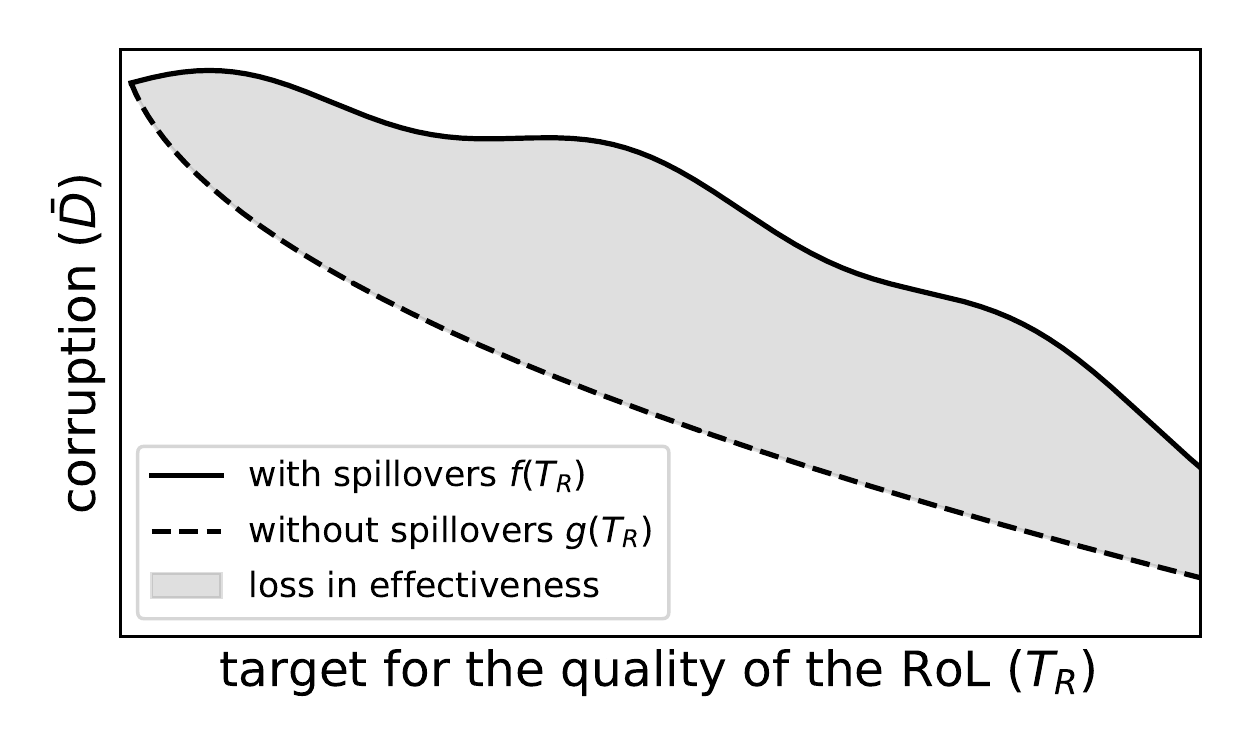}
    \label{fig:example}
\end{figure}

$L$ measures the area between the normalized versions of $f(T_R)$ and $g(T_R)$. Figure \ref{fig:example} provides a hypothetical example in which removing the spillovers generates a stronger $T_R-\bar{D}$ relationship. Then, the shaded area between both lines denotes the loss in effectiveness generated through a systemic component: the decentralized formation of corruption norms.

Relationships $f(T_R)$ and $g(T_R)$ are estimated through splines that interpolate values in the intervals $\{ \bar{D}_{1}, \dots , \bar{D}_{S} \}$ produced by the exogenous variation of $T_R$. This method facilitates an analytic computation of $L$ (discrete approximations yield similar results). In total, we specify $S=31$ (30 alternative values in addition to the initial estimation) for each country.

\subsection{Hypotheses}

Finally, and before proceeding to present our results, let us be more specific about the hypotheses to be tested. As we mentioned in section \ref{sec:intro}, we evaluate four families of hypotheses.

\subsubsection{Loss of effectiveness}
These are the main hypotheses of the paper. They provide evidence of a never-before measured effect caused by a systemic feature of corruption.

\begin{subhyp}
\begin{hyp}
Spillovers cause a loss in effectiveness of the RoL
\end{hyp}

\begin{hyp}
Losses in effectiveness of the RoL due to spillovers vary across countries
\end{hyp}
\end{subhyp}

\subsubsection{Policy profile}
These hypotheses take us beyond the current academic discussion of the RoL and corruption in order to shed some light on the potential outcomes of simultaneously implementing diverse policy prescriptions.

\begin{subhyp}
\begin{hyp}
Policy profiles that improve the RoL have a negative association with corruption
\end{hyp}

\begin{hyp}
The general improvement of policy profiles has a negative association with corruption 
\end{hyp}
\end{subhyp}

\subsubsection{Complementarity}
This hypothesis allows us to understand how, through spillovers, parallel prescriptions can boost reductions in corruption. These results can be extremely useful for rethinking policy evaluation and prescription in a systemic fashion.

\begin{hyp}
Policy issues with a negative association to corruption exhibit a stronger relationship when they have spillovers to/from the RoL, i.e. they are complementary in abating the diversion of public funds.
\end{hyp}

\subsubsection{Priority}
This set of hypotheses explores prescriptions packages associated to reductions in corruption. More specifically, they seek to answer: do effective prescriptions imply prioritizing the RoL over other topics?

\begin{subhyp}
\begin{hyp}
Reductions in corruption through improvements to the RoL are associated to higher relative target priorities
\end{hyp}

\begin{hyp}
Reductions in corruption through improvements to the RoL are associated to higher relative allocative priorities
\end{hyp}
\end{subhyp}

\section{Results 1: exogenous reform of the rule of law}\label{sec:results}

Besides testing the specific hypotheses from family I, our first group of results allow us to examine two broader and important questions about the public governance agenda: (1) for which type of countries do reforms to the RoL are most effective in curtailing corruption?; and (2) which countries experience heavier losses of effectiveness due to spillovers between public policies?

\subsection{Effectiveness of the rule of law}

\begin{figure}[ht]
    \centering
    \caption{Corruption as a function of the rule of law target}
    \includegraphics[scale=.5]{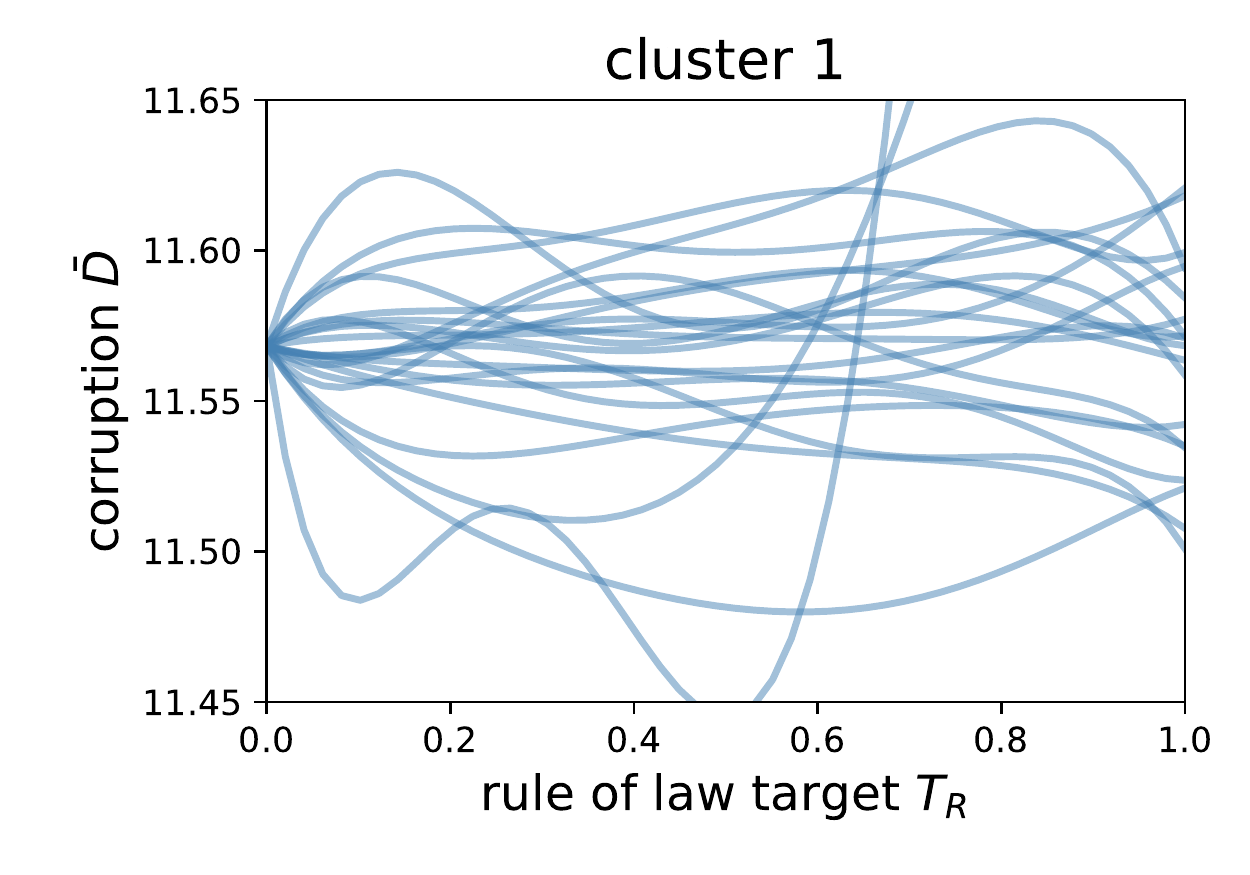}
    \includegraphics[scale=.5]{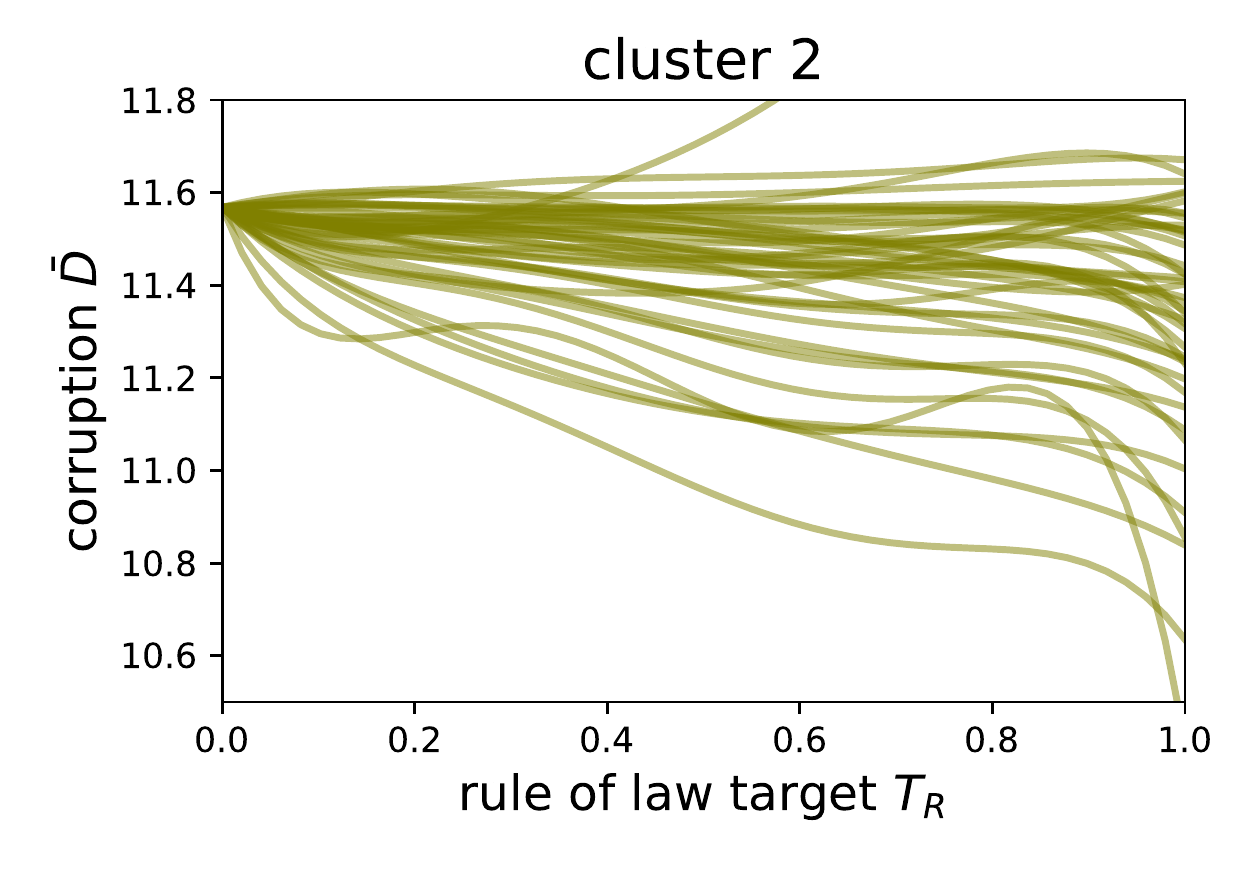}
    \includegraphics[scale=.5]{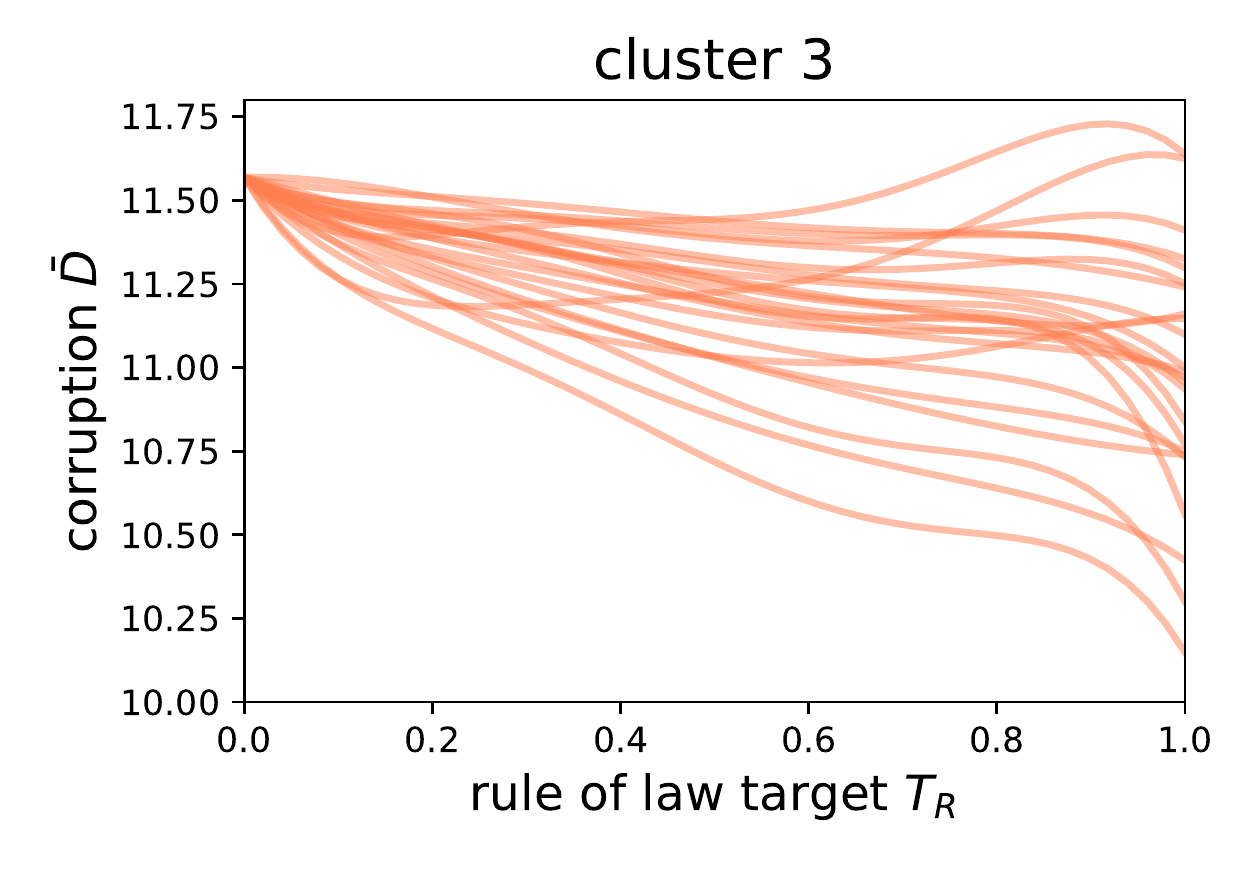}
    \includegraphics[scale=.5]{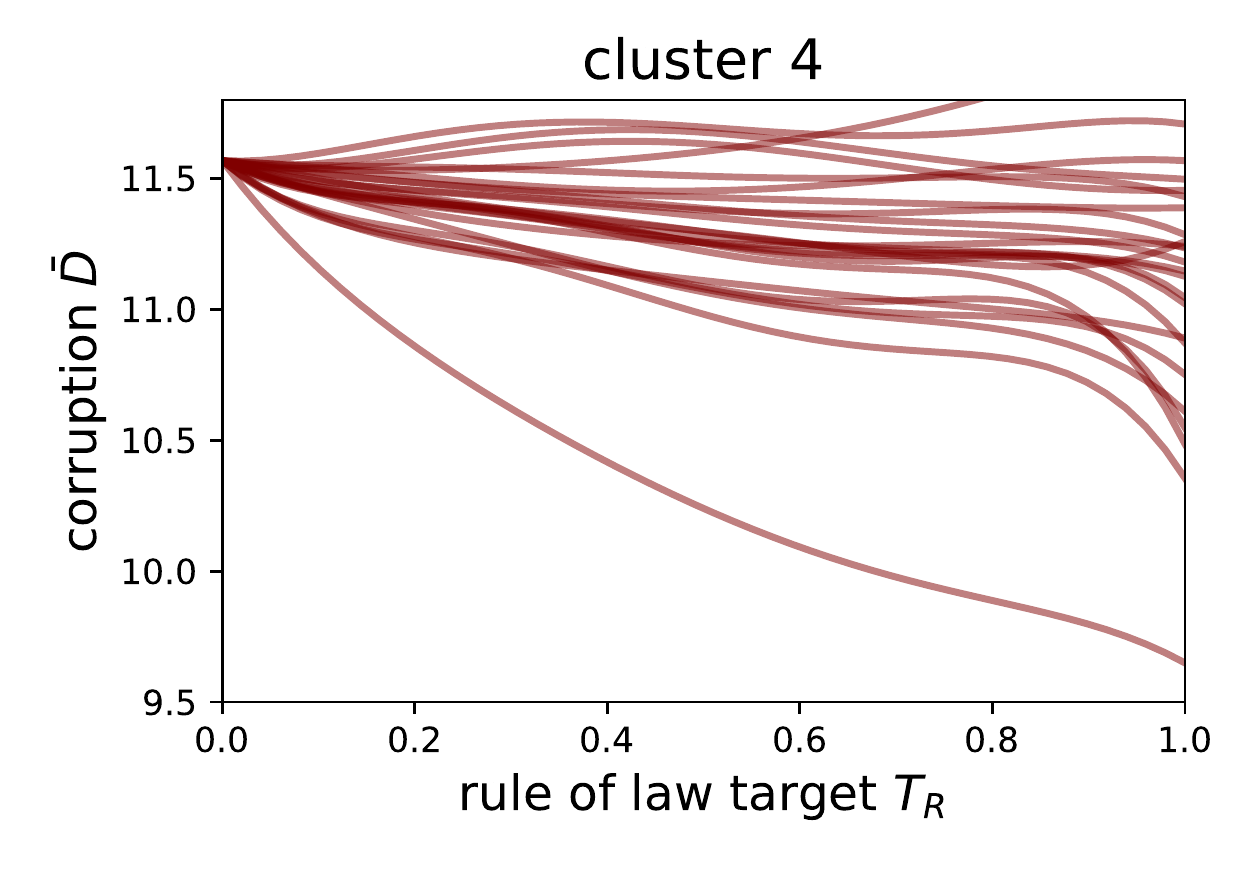}
    \caption*{Each line corresponds to the estimated effect of the RoL on corruption in a particular country. Each line was obtained from a polynomial spline interpolation of the intervals $\{ T_{R_1} \dots T_{R_S} \}$. For a visual comparison of their `slopes', all the splines have been vertically shifted to start with the same level of corruption. With the same purpose, the values in the horizontal axis have been normalized between 0 and 1. Clusters: (1) high, (2) middle-high, (3) middle-low and (4) low income.}
    \label{fig:cases}
\end{figure}

For the purpose of a visual comparison, we have grouped our estimations into the four clusters and normalized the horizontal axis so that $T_{R_1}=0$ and $T_{R_S}=1$. This facilitates the visual comparison of the $T_{R}-\bar{D}$ relationship across countries with very different initial conditions (\emph{i.e.}, controlling for the variation in the initial level of the RoL). Figure \ref{fig:cases} shows the $T_{R}-\bar{D}$ relationship for each country in the sample.

Figure \ref{fig:cases} shows several distinctive results. First, the RoL has a negative effect on corruption across many countries from clusters 2--4, and in very few from cluster 1 (the most advanced ones). This is consistent with the econometric literature, where the negative slope is mainly determined by the influence of developing countries. Second, in all clusters, there are several countries for which the RoL is not effective (the spline is flat); although, this phenomenon is more commonly observed in clusters 1 and 2. A tentative explanation is that more advanced nations have less room of improvement for the RoL, so the lack of variation in $\{ T_{R_1} \dots T_{R_S} \}$ generates weaker $T_{R}-\bar{D}$ relationships.\footnote{An alternative is that the principal-agent mechanisms dominate in these countries because effective monitoring and low impunity disincentivize corruption through frequent punishments and reductions of opportunities for diversions through effective reallocations. Under this view, it would not be surprising if the principal-agent-only view was majorly motivated from the experience of developed economies.} Third, even if the incidence is negative, it can be highly non-linear. Fourth, there are very few cases where the relationship is positive. Presumably, in these cases, the incentive effect is very small and substantially masked by the spillovers, so that the negative relationship disappears. At the same time, the allocative effect precludes assigning resources to policy issues that need them the most (\emph{i.e.} with the largest target gaps) which, in turn, produces a larger number of corruption events.    

\begin{figure}[ht]
    \centering
    \caption{Effectiveness of the RoL}
    \includegraphics[scale=.5]{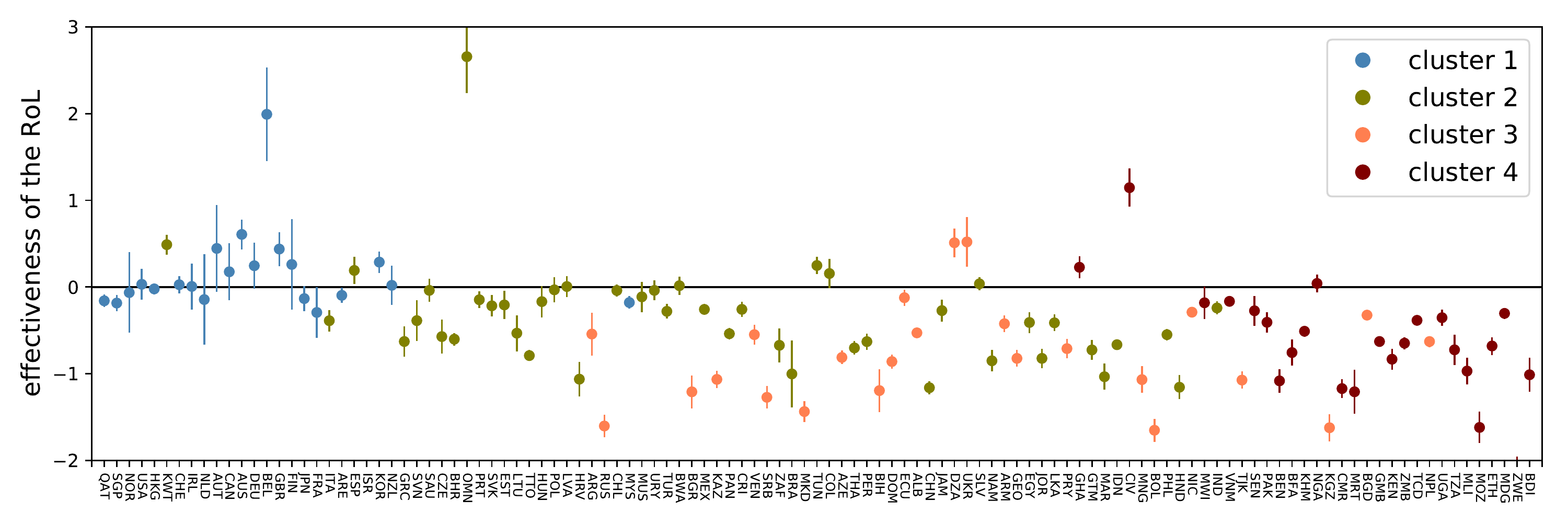}
    \caption*{$\beta$ coefficients estimated via OLS for the model $\bar{D} = \alpha + \beta T_R + \epsilon$. Countries in the horizontal axis have been sorted according to income per capita. The vertical lines correspond to 95\% confidence intervals (wider intervals show similar results).}
    \label{fig:slopes}
\end{figure}

As an exercise to confirm general results obtained in previous studies, let us estimate the relationship $f(T_R)$ for each country via linear regression. That is, we estimate the model $\bar{D} = \alpha + \beta T_R + \epsilon$. If the $\beta$ coefficient results negative and significant in most countries, it means that our model generates data that is consistent with the empirical literature. Out of 115 countries in the sample, 79 (69\%) present a negative $\beta$ coefficient. Interestingly, and also consistent with previous findings, a negative and significant association between corruption and the RoL exists mainly in developing countries, not among the industrialized nations. For instance, in cluster 1, countries with a negative and significant $\beta$ represent only 23\%; in cluster 2 they are  68\%; in clusters 3 they are 92\%; and cluster 4 has 88\%. We provide a more detailed picture across countries in Figure \ref{fig:slopes}, where nations have been sorted by income per capita. Clearly, the less developed a country is (orange and dark-red dots), the better the chances to reduce corruption via improvements to the RoL.\footnote{Note that there are some country-cases where the estimated average relationship is positive, specially for advanced nations.} These results confirm, too, our scepticism on modelling frameworks that assume parameter homogeneity across countries.

\subsection{Loss of effectiveness}

We have argued that the mismatch between optimistic predictions and poor outcomes from improvements to the RoL can be explained by a generalized indifference toward the systemic features of corruption. Here, we focus on one such feature --spillovers-- and provide novel estimates on the $T_R-\bar{D}$ relationship that suggest a loss of effectiveness of the RoL.

\begin{figure}[ht]
    \centering
    \caption{Change in the $T_R-\bar{D}$ relation from spillover effects}
    \includegraphics[scale=.45]{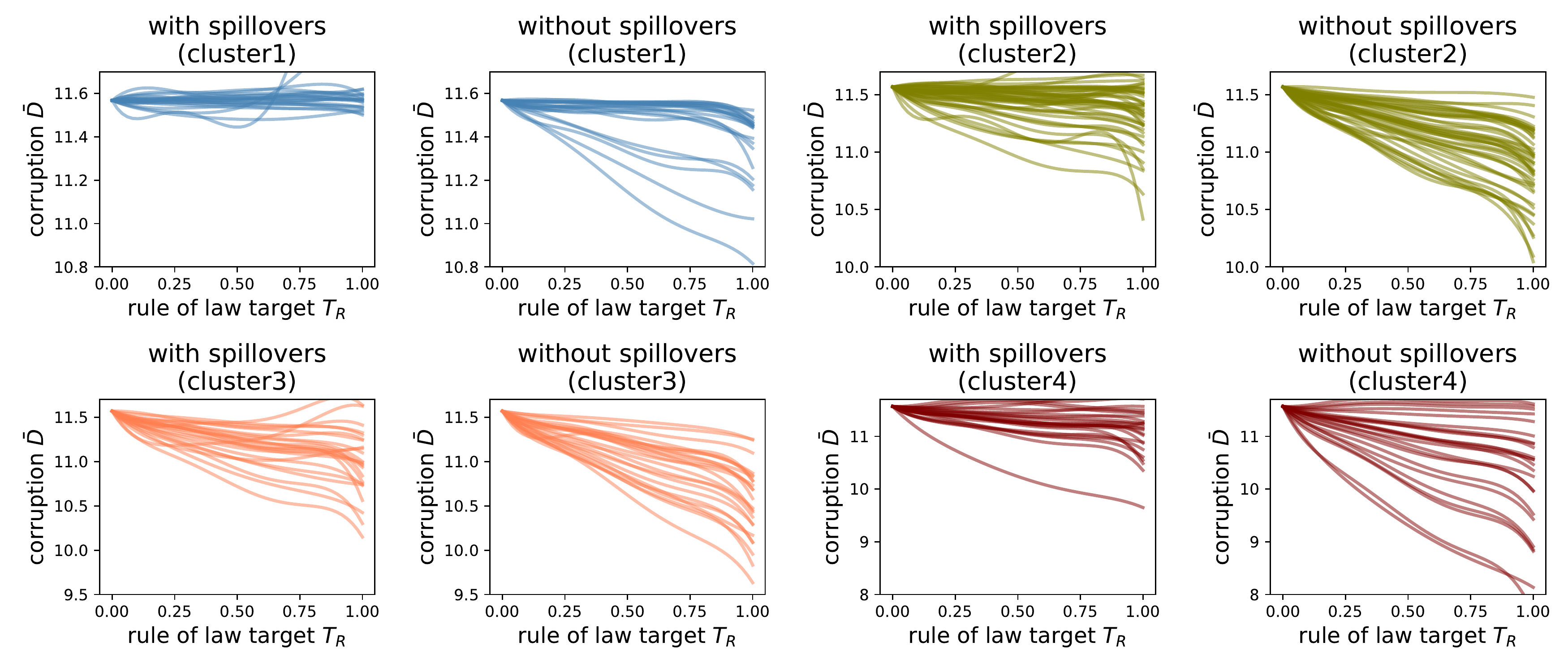}
    \caption*{Splines of the $T_R-\bar{D}$ relation by cluster and model specification (with and without spillovers). For each cluster, the left panel shows the splines estimated for the model without spillover effects, while the right panel corresponds to the splines from Figure \ref{fig:cases}}
    \label{fig:network_effect}
\end{figure}

For a qualitative comparison, Figure \ref{fig:network_effect} contrasts the estimations from Figure \ref{fig:cases} (the left panels of each cluster) against the ones obtained by removing the spillover network (the right ones; here $\mathbb{A}$ is an identity matrix). Just from eyeballing the plots, it is clear that the negative incidence of the RoL on aggregate corruption becomes stronger without spillovers. This suggests that the systemic feature of spillover effects dampens the effectiveness of the RoL. Note that, without spillovers, almost all countries present a negative relationship and that, only after introducing a network, we obtain that the countries where the RoL can be effective are the developing ones. Qualitatively, this outcome is similar to the one obtained when moving from pooled to stratified regressions. However, we have achieved this through a micro-funded model, allowing us to produce country-level estimates. Furthermore, country-level losses in effectiveness are consistent with the poorly-realized real-world outcomes from reforms to the RoL experienced by governments and multilateral organizations.

\begin{figure}[ht]
    \centering
    \caption{Examples of `similar' countries with different losses in the effectiveness of the RoL}
    \includegraphics[scale=.4]{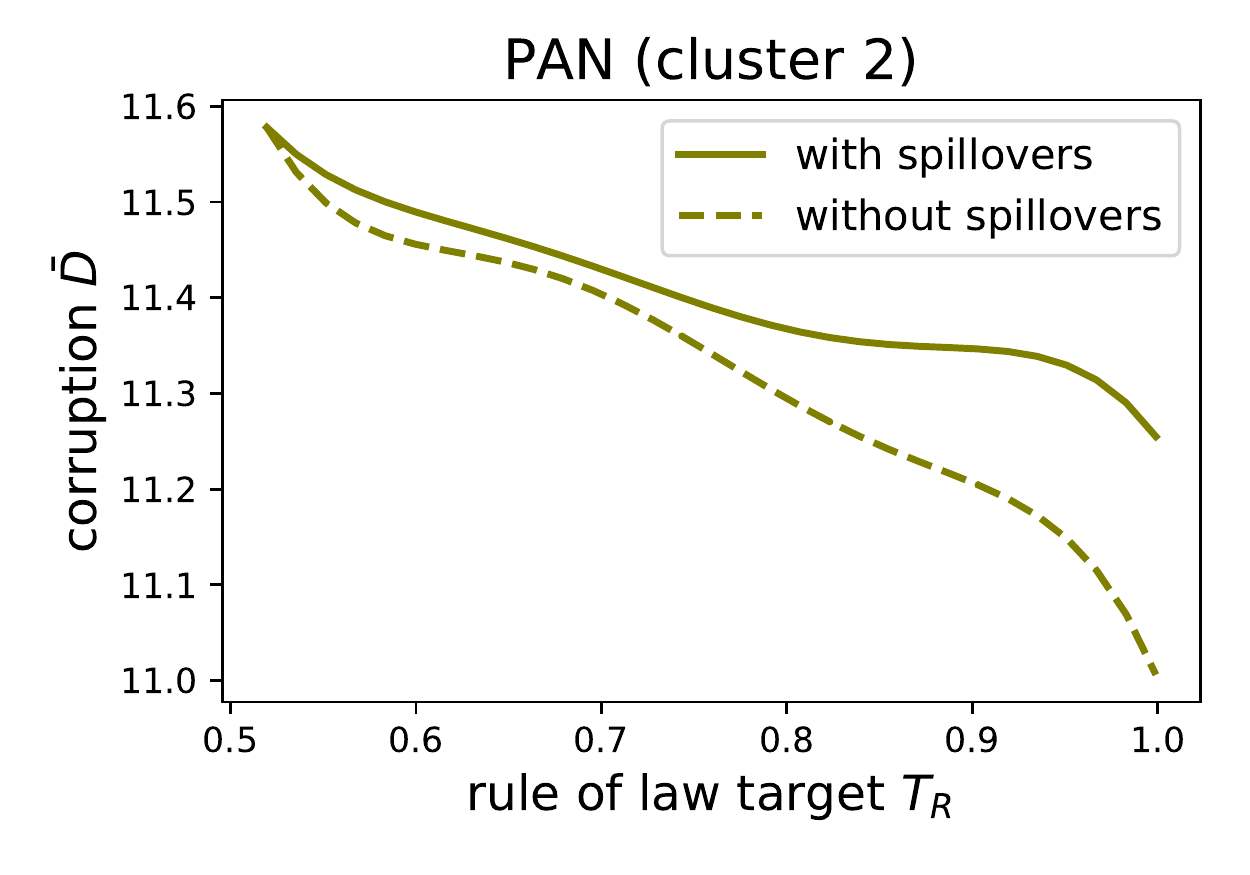}
    \includegraphics[scale=.4]{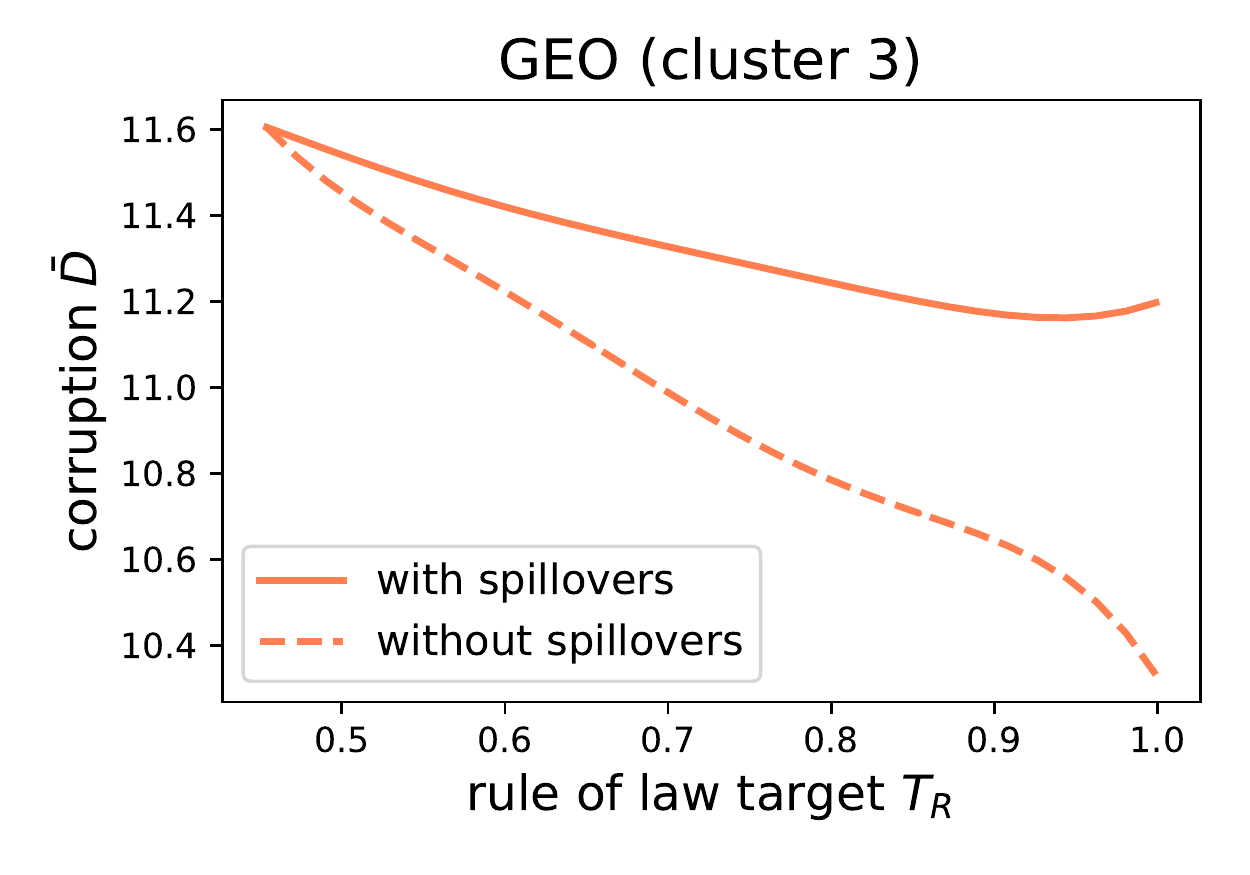}
    \includegraphics[scale=.4]{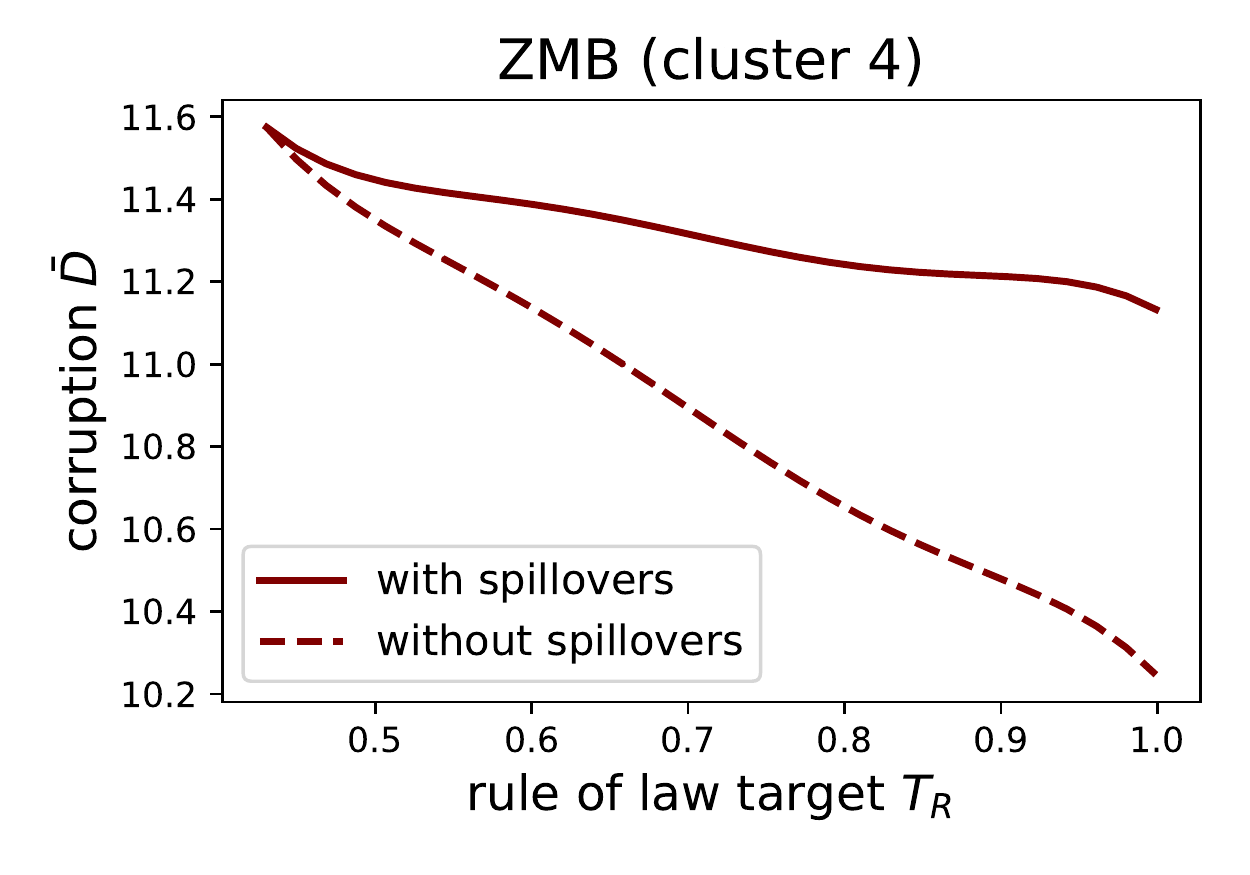}
    \includegraphics[scale=.4]{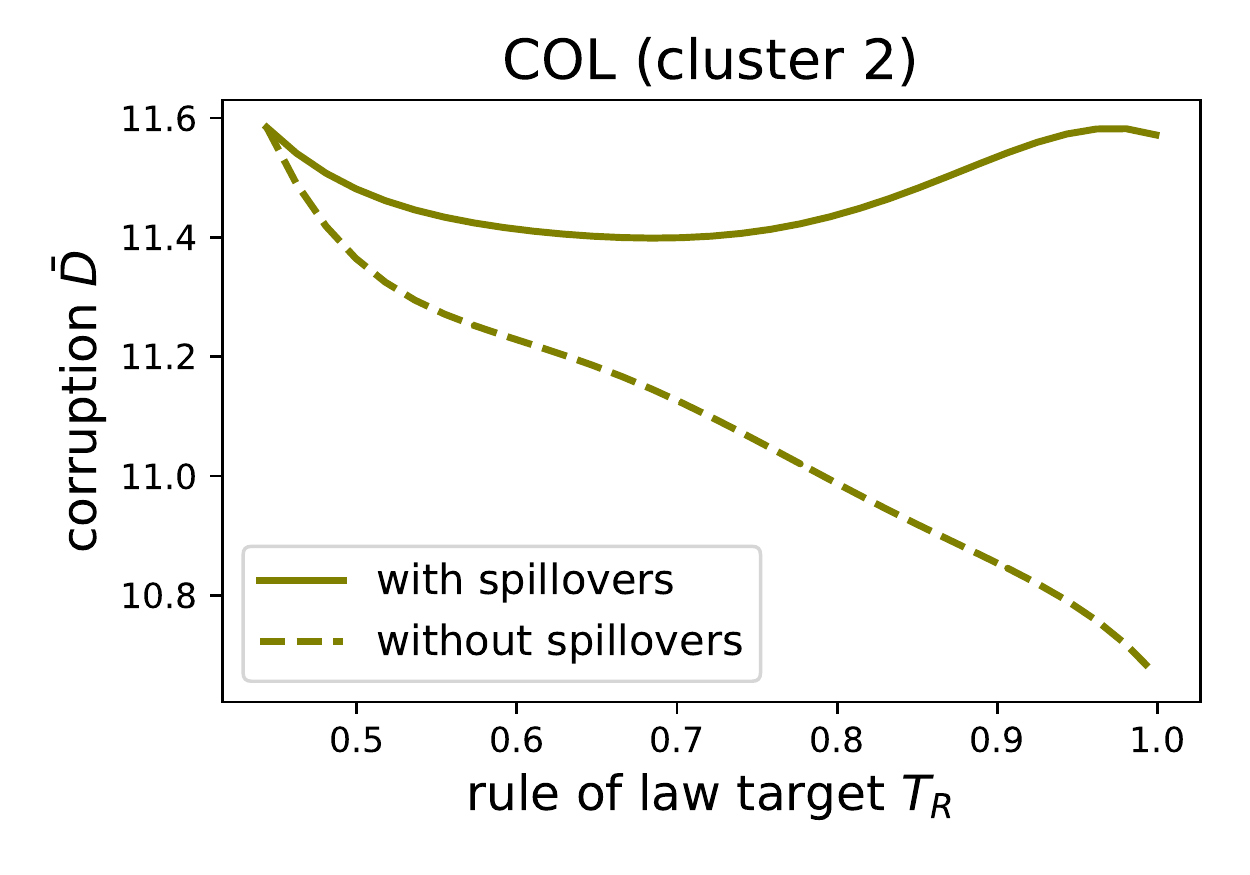}
    \includegraphics[scale=.4]{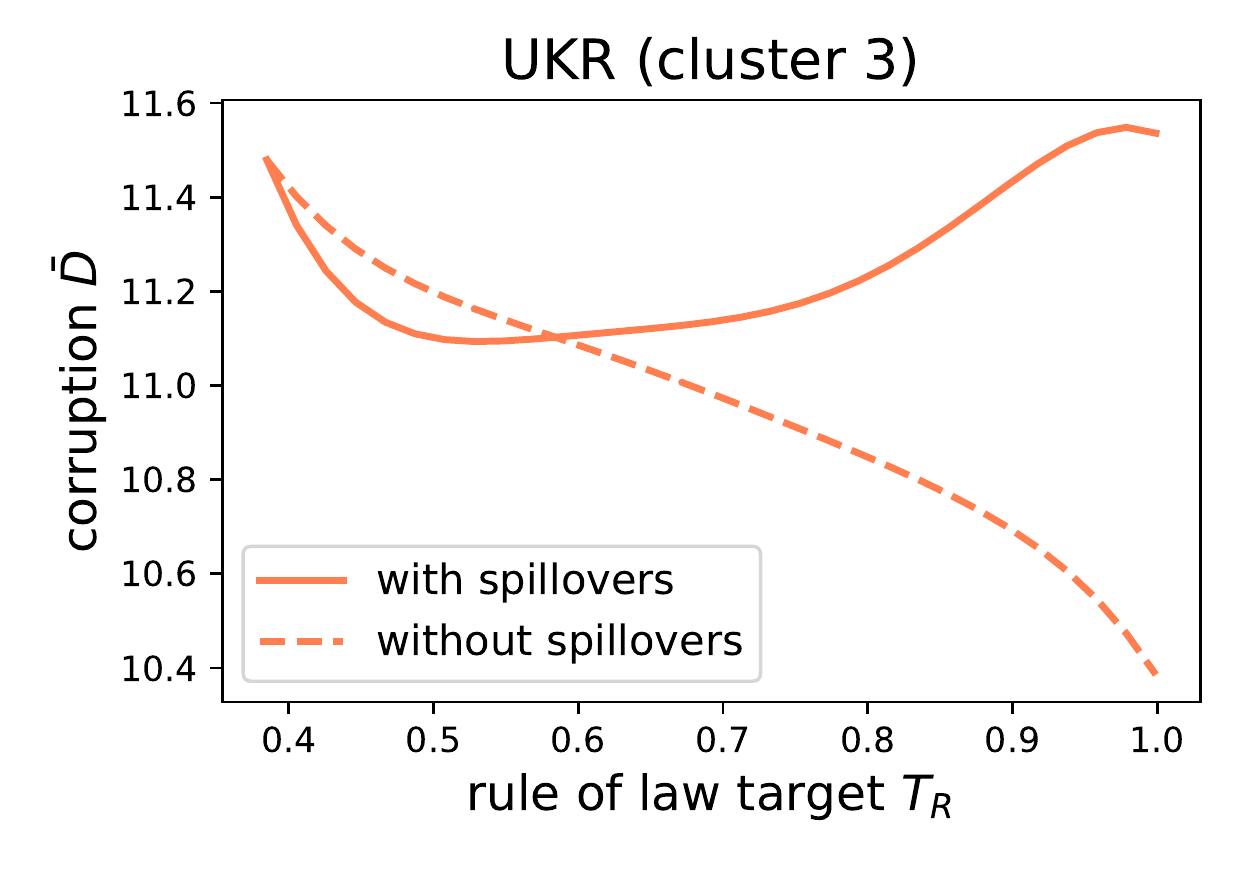}
    \includegraphics[scale=.4]{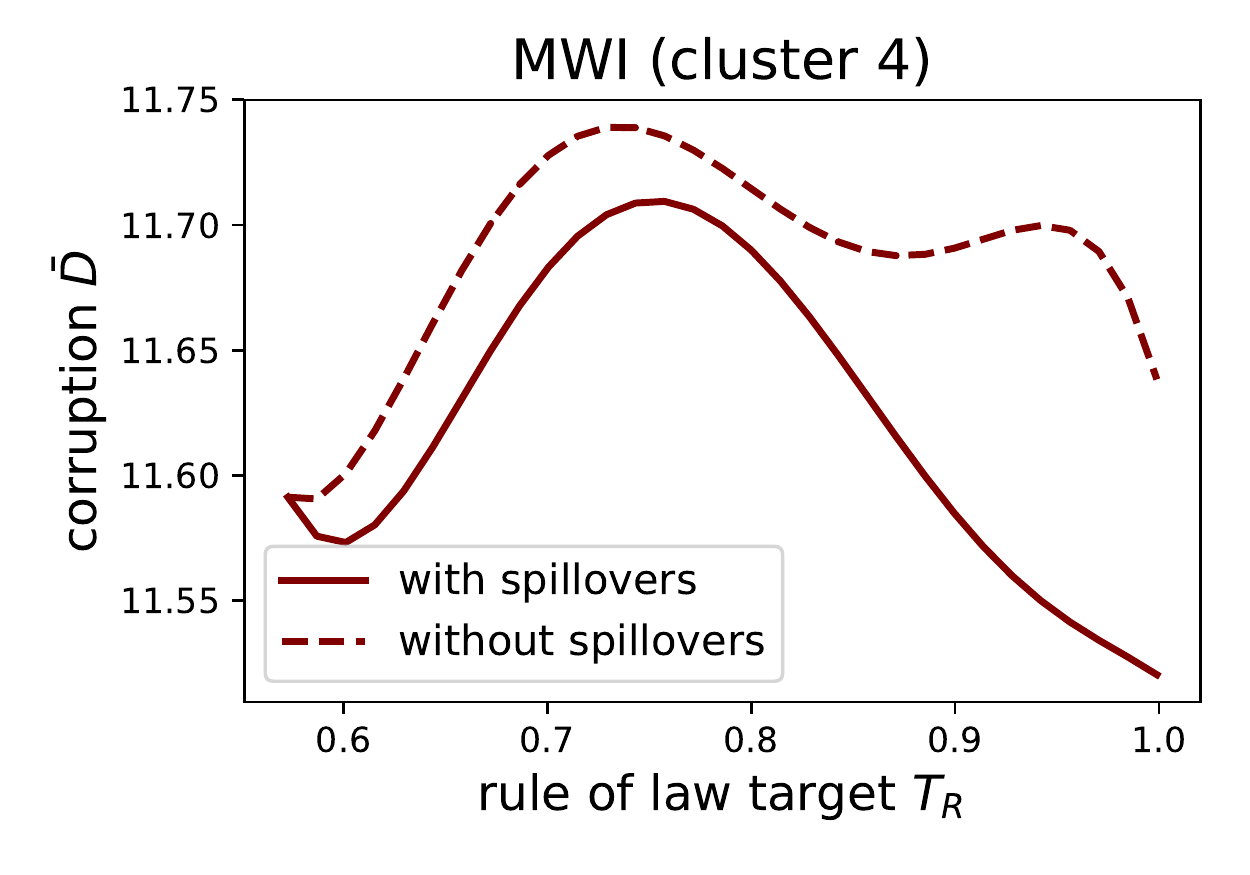}
    \caption*{Two countries from cluster 2, 3 and 4 have been chosen for illustrative purposes. Each pair consists of neighboring nations.}
    \label{fig:examples}
\end{figure}

Now, let us look in more detail at the changes in the $T_R-\bar{D}$ relationship caused by the spillover network. For this, we consider six countries from the least-developed clusters 2--4; two per each cluster.\footnote{These are the clusters where improvements to the RoL tend to be more effective in reducing corruption.} Furthermore each pair of countries within a cluster shares a geographical border. Hence, within-pair differences would make a strong case for the importance of context-specificity. Figure \ref{fig:examples} shows each country with its corresponding spline with and without spillovers. As we can see, countries from cluster 2 exhibit sharp differences in the loss of effectiveness. In fact, the RoL becomes relatively ineffective in Colombia, while in Panama --despite the spillovers-- it continues to be a useful policy instrument. The case of Ukraine, in cluster 3, exhibits a slight gain in efficiency for small improvements to the RoL but, after certain point, it shows losses. This country demonstrates how spillovers can generate strong non-linearities. In contrast, its neighbor Georgia shows a linear relation through most values of $T_R$, and an increasing loss in effectiveness. Finally, the pair from cluster 4 shows a well behaved case (Zambia) and a degenerate one (Malawi). In the latter, for all the range of $T_R$, there are gains in efficiency from the spillovers. Although this case is an outlier, it is useful to illustrate that it is possible to observe unlikely outcomes.\footnote{Note: (1) a positive relationship $T_R-\bar{D}$, even when controlling for systemic factors (increasing segment of the dashed-curve); and (2) incentive and allocative effects that impact corruption in the same direction as the systemic factors (decreasing segments of the curves).}

\begin{figure}[ht]
    \centering
    \caption{Country-level index of the loss of effectiveness to the RoL}
    \includegraphics[scale=.5]{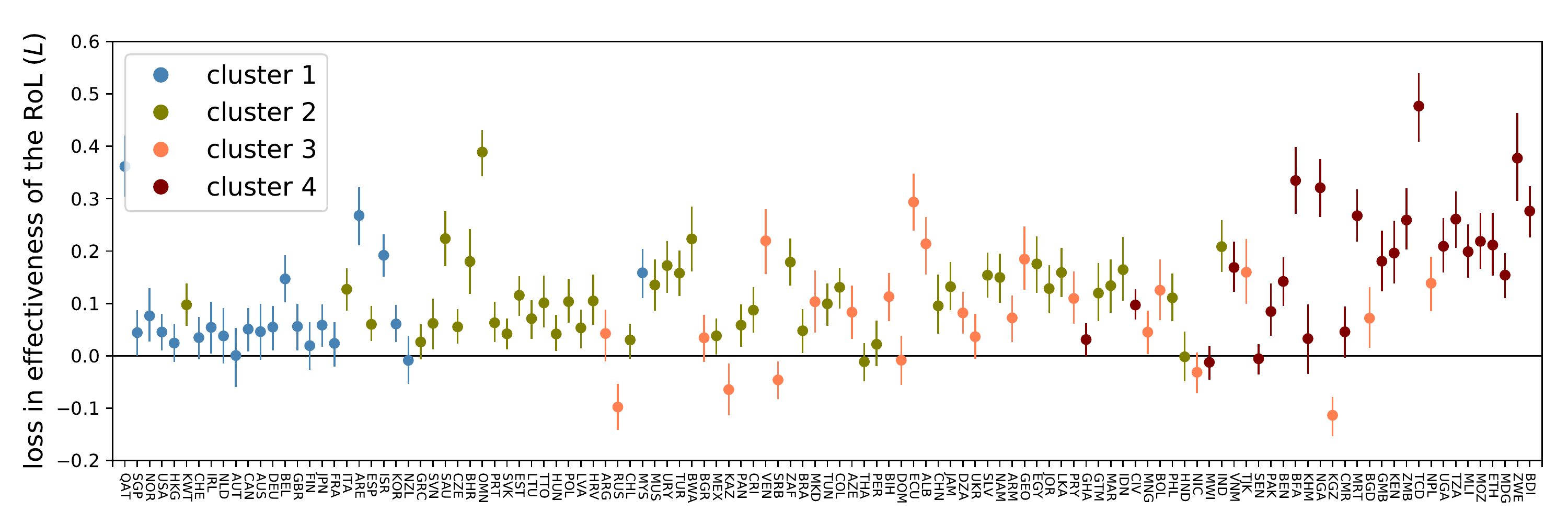}
    \caption*{A positive index denotes loss of effectiveness while a negative value signifies a gain. The vertical lines denote the 95\% bootstrap confidence interval (wider intervals show similar results).}
    \label{fig:indices}
\end{figure}

In order to test hypotheses I.a and I.b, we compute the $L$-index introduced in equation \ref{eq:index}. Recall that a positive index denotes a loss in effectiveness of the RoL, while a negative one indicates gains. In support of hypothesis I.a, we find that the loss in effectiveness is statistically significant in the majority of the countries (77\% of the sample). When looking at the cross-national variation, we find that only 64\% of cluster 1 experience a significant loss, but these represent 89\% in cluster 2, 63\% in cluster 3 and 83\% in cluster 4. Such a striking difference in loss of effectiveness between developed and developing countries highlights the importance of the context-specific systemic features of corruption and supports hypothesis I.b. Figure \ref{fig:indices} provides a detailed picture of these findings.

Combining the results from Figures \ref{fig:slopes} and \ref{fig:indices}, we conclude that (1) there are several countries from cluster 4 where reforms to the RoL are effective; (2) their impact could be significantly strengthened through strategies that consider systemic factors; (3) for some countries in clusters 2 and 3, the RoL will not be effective unless systemic elements are considered; (4) for some countries in clusters 1 and 2, reforms to the RoL will not be able to abate aggregate corruption.  These results coincide with the observed failure in the implementation of the governance agenda in many developing countries. Clearly, any attempt to avoid this loss in effectiveness would require profiles of coordinated policies that exploit the systemic nature of corruption.

\section{Results 2: multidimensional policy profiles}\label{sec:policy}

So far, we have studied the effect that isolated changes to the target of the RoL have on corruption. These changes, however, are rarely observed in the real world. This is so because, when governments set development strategies, they establish simultaneous reforms across different dimensions. Take, for instance, the example of a government receiving advise from experts on different areas (\emph{e.g.}, macroeconomic policy, public health, telecommunications, etc.). By collecting such recommendations, multidimensional policy profiles are assembled and, eventually, policy priorities established. Thus, it is natural to hypothesize whether the systemic features of corruption play a role in the effectiveness of the RoL as part of a policy profile. For example, if the topology of the spillover network facilitates complementarities between the RoL and other policy issues, one would expect a stronger incidence on corruption. It follows, thus, that it is important to understand the role of the RoL under different policy profiles.

\subsection{Policy profiles and the rule of law}

First, we are interested in testing the association between improvements in policy profiles and corruption. We do it by testing hypotheses II.a and II.b. For this, it is necessary to introduce exogenous variation not only in $T_R$ but in the entire target vector $T$. More specifically, for each policy issue $i$, $T_i$ is created by randomly sampling the interval $(I_{i,n}, 1)$ under a uniform distribution (where $n$ is the last period in the sample). In this way, we generate random uncorrelated policy profiles that prescribe improvements across all policy issues.

\begin{figure}[ht]
    \centering
    \caption{The RoL and corruption across policy profiles}
    \includegraphics[scale=.4]{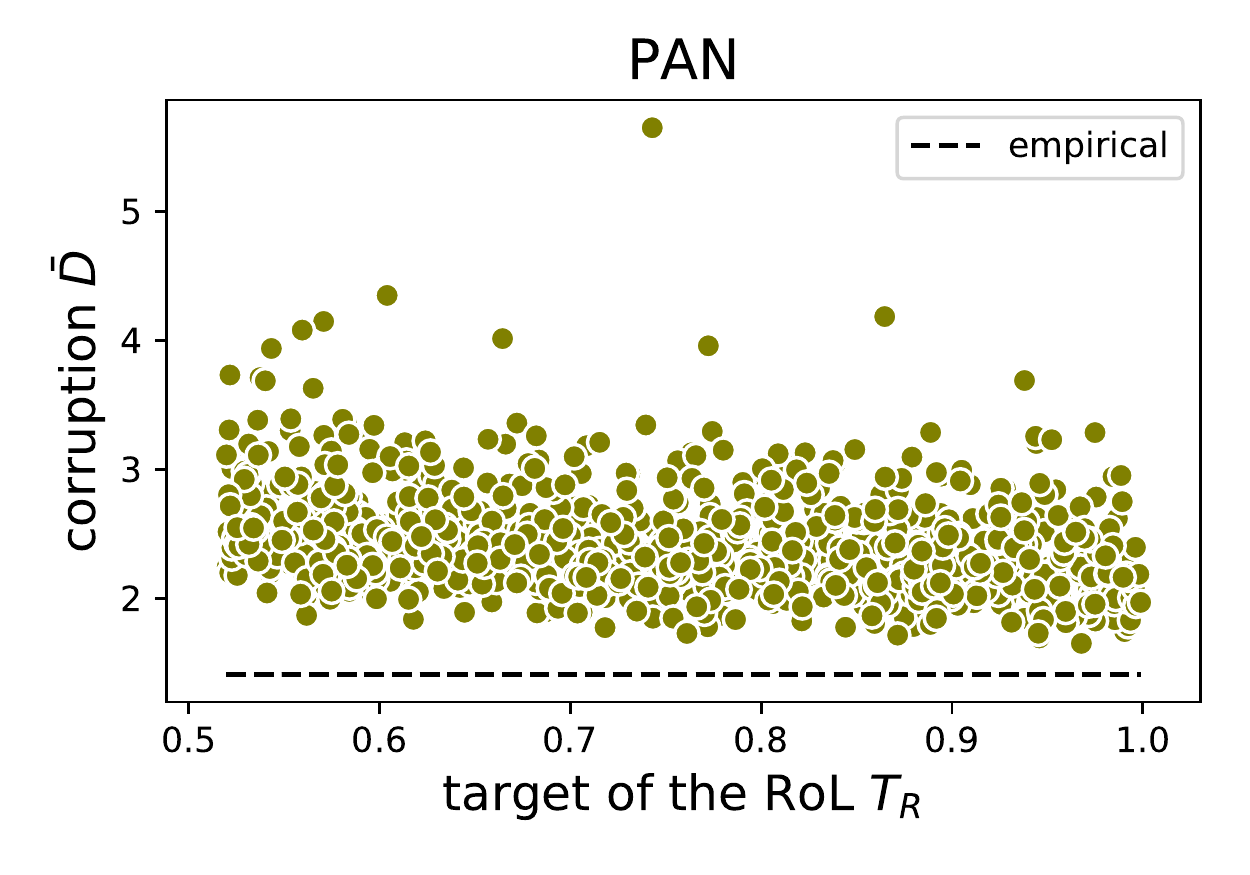}
    \includegraphics[scale=.4]{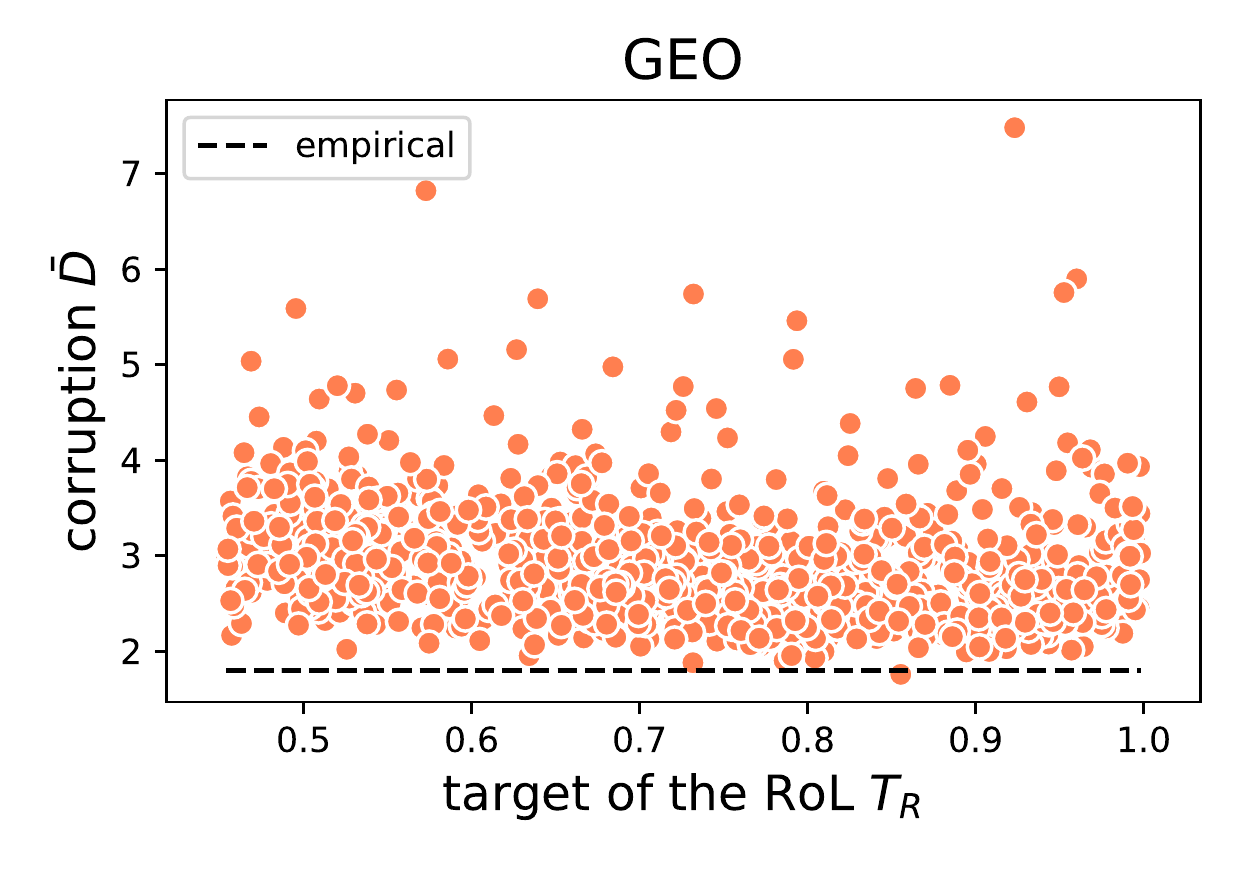}
    \includegraphics[scale=.4]{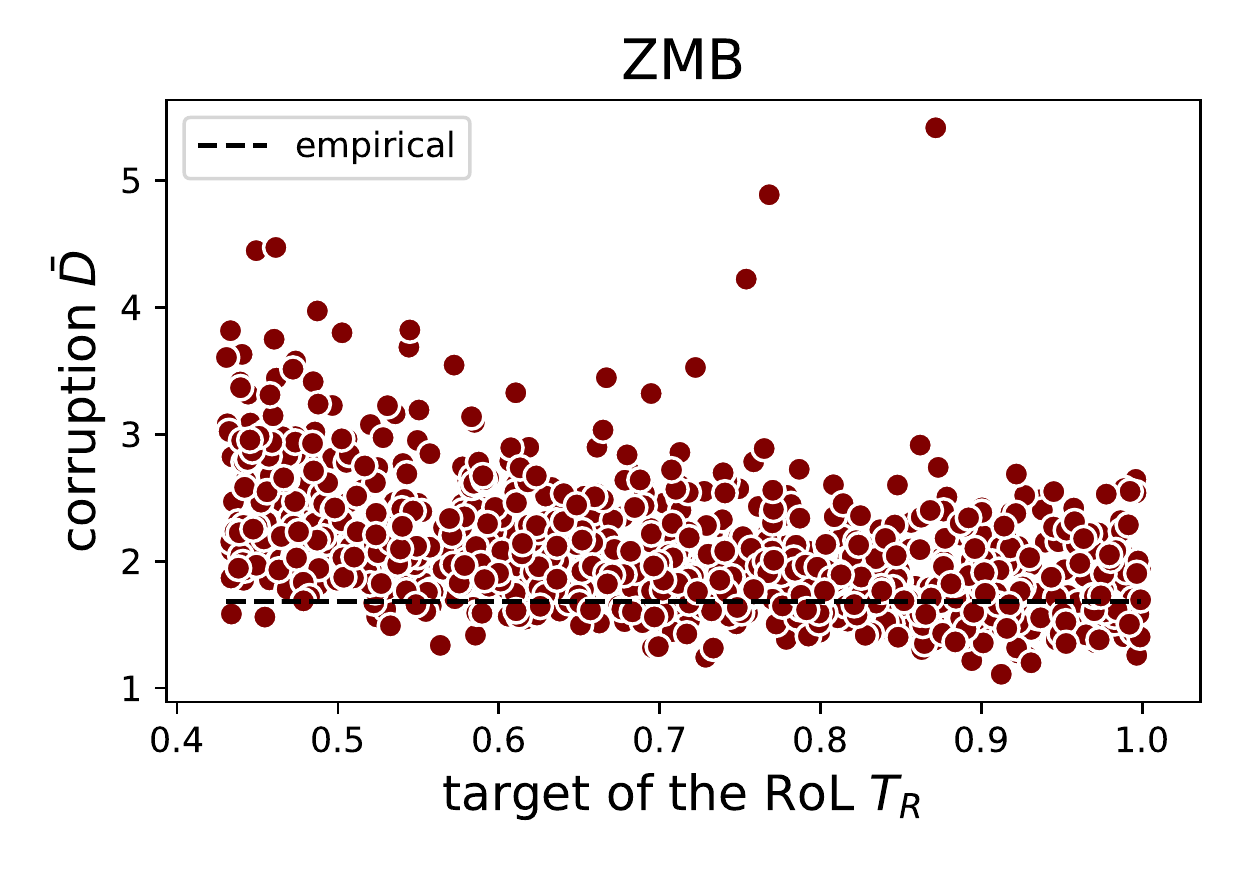}
    \includegraphics[scale=.4]{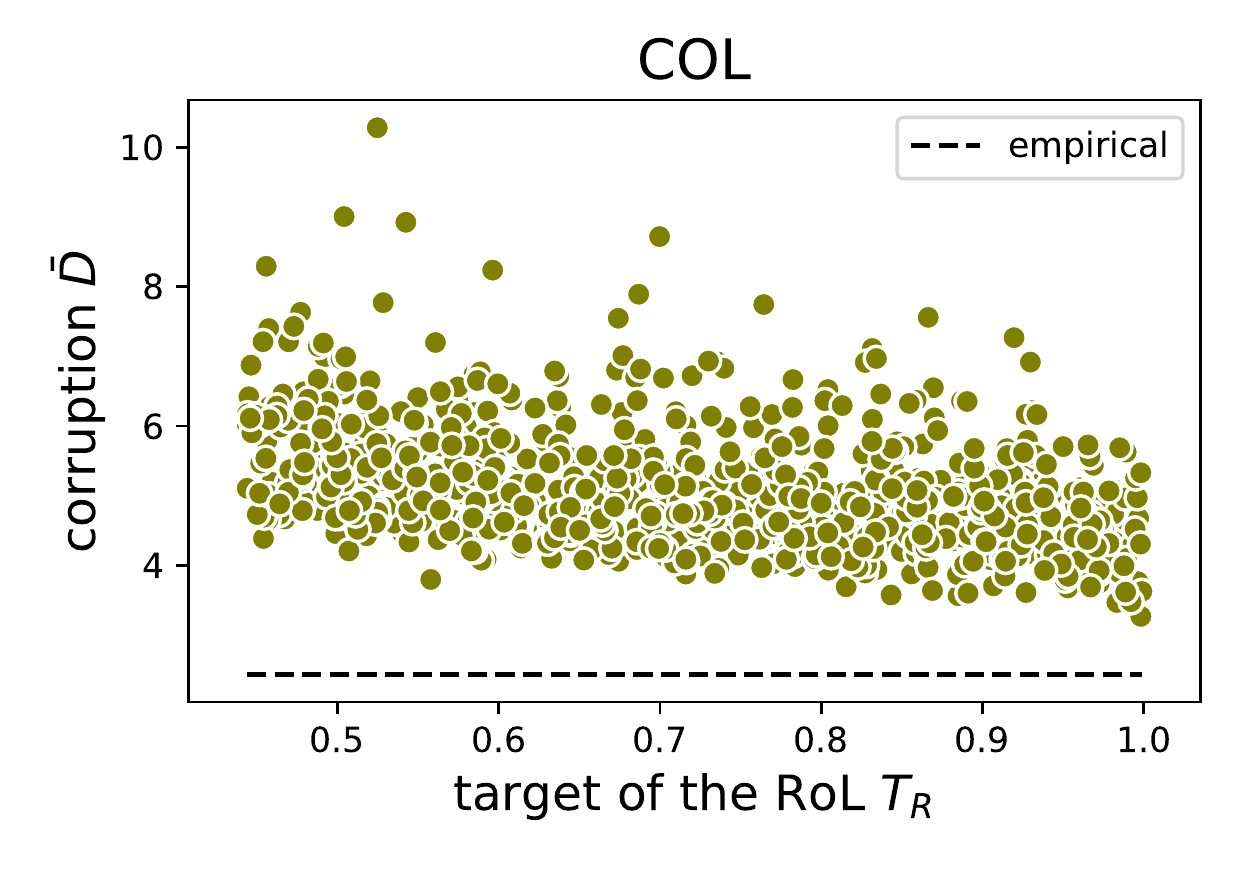}
    \includegraphics[scale=.4]{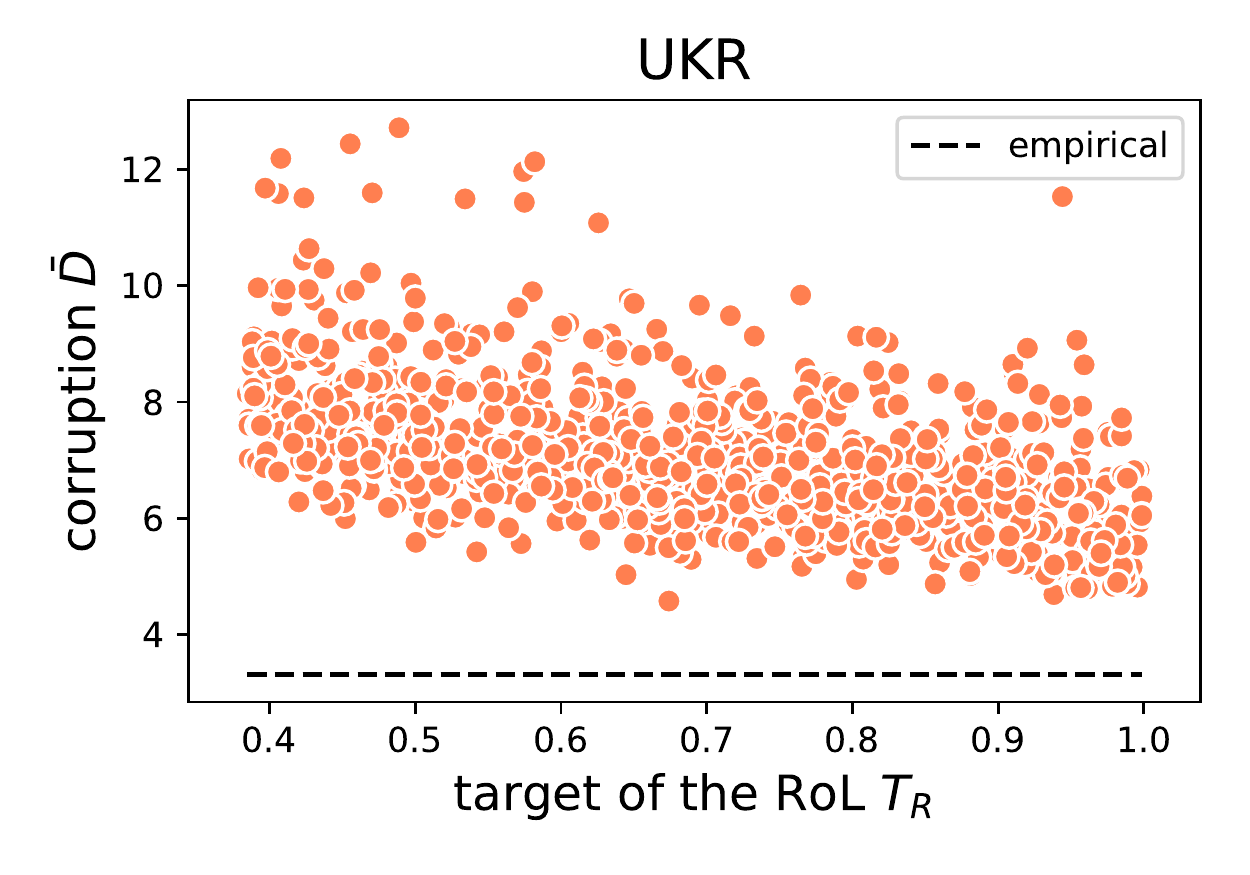}
    \includegraphics[scale=.4]{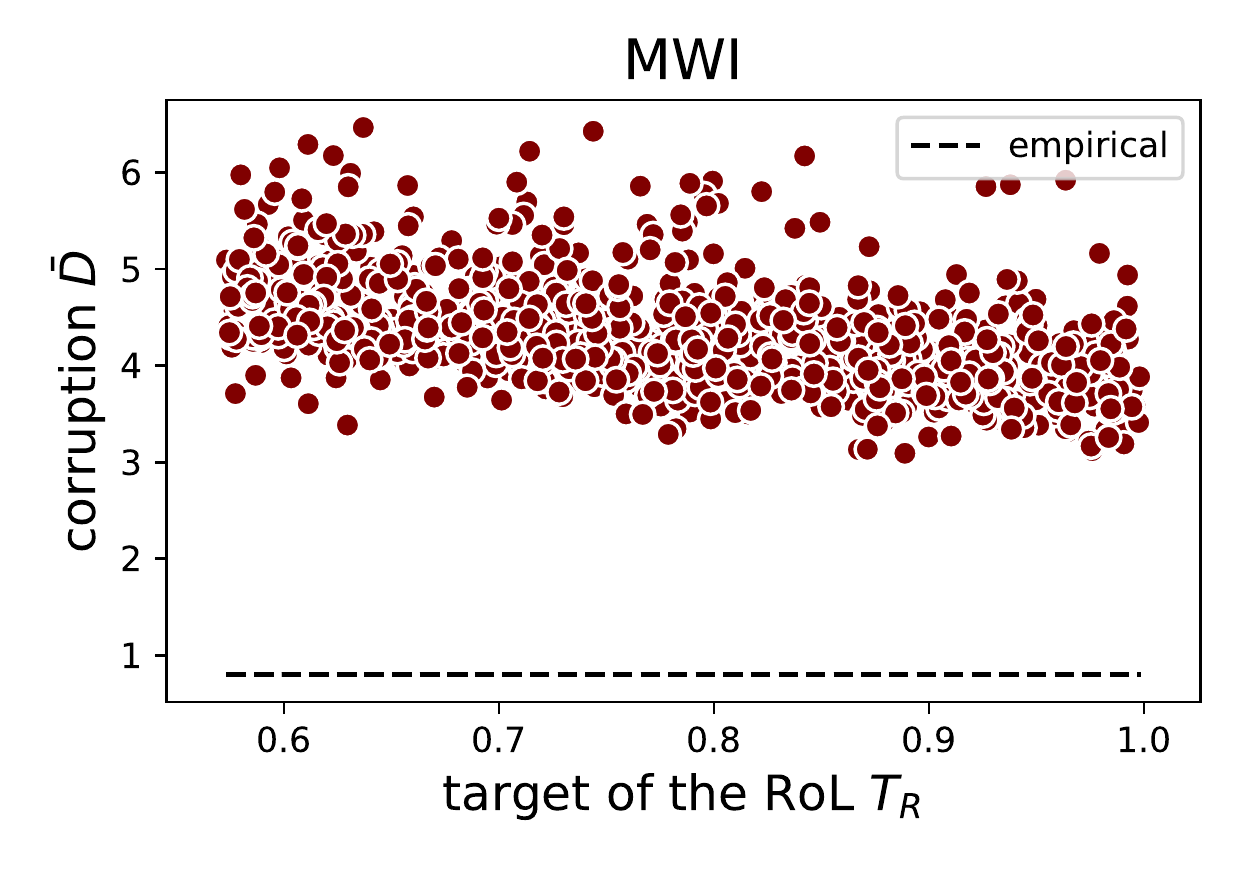}
    \caption*{Two countries from cluster 2, 3 and 4 have been chosen for illustrative purposes. Each pair consists of neighboring nations. Each policy profile was generated by sampling target levels $T_i \sim U(I_{i,n}, 1)$.}
    \label{fig:rol_pack}
\end{figure}

For illustration purposes, let us consider the six country cases presented in Figure \ref{fig:example}. Each panel in Figure \ref{fig:rol_pack} shows the relationship between $T_R$ and $\bar{D}$ through 1000 different policy profiles. Visually, all countries seem to present a negative relationship between the target of the RoL and diversion of public funds, suggesting a negative association between the RoL and aggregate corruption, unconditional on policy profile. This association, however, varies from country to country. For example Georgia appears to have the weakest association while Colombia or Malawi the strongest.\footnote{Note that the empirical level of corruption seems to be lower than the one generated through random policy profiles. This result is expected since all random profiles present targets above the observed levels and, hence, reaching them requires more corruption events.} 

\begin{table}[ht]
\centering
\caption{Average correlation between $T_R$ and $\bar{D}$}
\begin{center}
\begin{tabular}{ l r r r r r}
\hline
\hline
Statistic & Pooled & Cluster 1 & Cluster 2 & Cluster 3 & Cluster 4\\
\hline
mean Spearman ($T_R$) & -0.31 & -0.01 & -0.3 & -0.47 & -0.46\\
mean $p$-value ($T_R$) & 0.03 & 0.18 & 0.0 & 0.0 & 0.0\\
negative \& significant (\%) & 88.7 & 36.36 & 97.78 & 100.0 & 100.0\\
\hline
mean Spearman ($\bar{T}$) & 0.08 & 0.2 & 0.08 & 0.0 & 0.03\\
mean $p$-value ($\bar{T}$) & 0.17 & 0.0 & 0.15 & 0.0 & 0.21\\
negative \& significant (\%) & 2.61 & 0.0 & 0.0 & 4.17 & 4.17\\
\hline
$N$ & 115 & 22 & 45 & 24 & 24\\
\hline
\end{tabular}
\end{center}
\caption*{$N$ indicates the number of countries included in the analysis. Statistical significance is at 95\% (wider intervals show similar results).}
\label{tab:correlation}
\end{table}

In order to test hypotheses II.a and II.b, we compute the Spearman correlation between $T_R$ and $\bar{D}$ for each country. The first three rows in Table \ref{tab:correlation} show the average correlation and mean $p$-values for the entire sample and by cluster. In addition, they show the percentage of countries in each group that have a negative and significant correlation. Clearly, these results support hypothesis II.a on a negative association between the RoL and aggregate corruption across policy profiles.

The role of the RoL in reducing corruption through policy profiles becomes clearer when testing hypothesis II.b. For this, we calculate the average value $\bar{T}$ of the entire target vector and its correlation with $\bar{D}$. The intuition behind this exercise is to verify if improvements to arbitrarily chosen issues also reduce corruption or if they need to be accompanied by improvements to the RoL. The bottom set of rows in Table \ref{tab:correlation} present strikingly different results from the top one. Clearly, they do not support hypothesis II.b and suggest that the RoL plays an important role regardless of the proposed policy profile. In a way, we can think of these results in the following way: in order to curve corruption, improvements to the RoL seem to be necessary but not sufficient.

\subsection{Complementarities for abating of corruption}

We have established the necessity of improving the RoL in order to fight corruption. At the same time, we have argued that, by itself, the RoL is insufficient to substantially ameliorate the problem. Thus, a natural follow-up question is whether there are other policy issues that can serve as complements to the RoL. Such complements could be extremely helpful to design better-articulated policy profiles.

Given the theory and the model presented in this paper, it make sense to look for complementary policy issues in the spillover network. Here, we can think about complementarities in two different ways. First, the presence of incoming links to the RoL implies that contributions to connected policy issues will help dampening corruption through indirect improvements in the quality of the law. Second, the existence of outgoing links from the RoL implies that enhancements in this indicator will produce spillover effects, reduce target gaps and, then, diminish the frequency of corruption. In order to evaluate these ideas, we test hypothesis III, which states that, from all the policy issues with a negative association to aggregate corruption, those that share links to/from the RoL have a stronger association. In order to test this, we use the simulated data on policy profiles to estimate the model

\begin{equation}
    \bar{D} = \beta_1T_1 + \dots + \beta_NT_N + \epsilon
\end{equation}
for each country. Note that, by construction, the independent variables are uncorrelated and their observations are independent because they are produced from separate simulations. Hence, having such a large number of covariates does not produce multicolinearity. 

The $\beta$ coefficients approximate the association between each indicator and $\bar{D}$. For a given country, we isolate those coefficients that are negative and significant (with a $p$-value $\leq 0.1$). Then, we separate them into those with a connection to the indicator of the RoL (incoming or outgoing) and those without it. Finally, we test whether the group of covariates with links have $\beta$ coefficients significantly different from those without. 

In order to fulfill the assumptions of a statistical test for related samples, we restrict the dataset to those countries that have at least one negative and significant coefficient with a link to the RoL and at least one without. We also remove the coefficient corresponding to \emph{control of corruption} because its corresponding indicator has an explicit effect in the model through equation \ref{eq:catch}, which would bias the results in favor of hypothesis III. For each country in the reduced sample, we compute the average $\bar{\beta} = \sum_i^n|\beta_i|/n$ in each group (the groups being: with and without link). Finally, we pair the $\bar{\beta}$ between groups by country (\emph{e.g.}, Colombian $\bar{\beta}$ with link \emph{versus} Colombian $\bar{\beta}$ without link).

Table \ref{tab:testDiff} presents some descriptive statistics from the sub-sample and the outcomes of non-parametric mean-equality tests for related samples. Note that the average $\beta$ coefficient is always higher among policy issues with links to the RoL than among those without them, supporting hypothesis III. This is robust even after dividing the sample into clusters. The results from the Wilcoxon sign-rank tests confirm that these differences are statistically significant.

\begin{table}[ht]
\centering
\caption{Test for differences between having and not having links to the RoL}
\begin{center}
\begin{tabular}{ l r r r r r}
\hline
\hline
Statistic & Pooled & Cluster 1 & Cluster 2 & Cluster 3 & Cluster 4\\
\hline
\hline
$\mu(\bar{\beta})$ link & 1.15 & 3.17 & 0.98 & 0.13 & 0.27\\
$\sigma(\bar{\beta})$ link & 1.85 & 2.9 & 1.12 & 0.03 & 0.18\\
$\mu(\bar{\beta})$ no link & 0.2 & 0.6 & 0.12 & 0.08 & 0.07\\
$\sigma(\bar{\beta})$ no link & 0.33 & 0.59 & 0.05 & 0.03 & 0.05\\
$N$ & 30 & 6 & 14 & 6 & 4\\
\hline
$W$-statistic & 9773.0 & 426.0 & 1414.0 & 1435.0 & 69.0\\
$p$-value & 0.0 & 0.0 & 0.0 & 0.0 & 0.0\\
\hline
\end{tabular}
\end{center}
\caption*{The $W$-statistic corresponds to the Wilcoxon sign-rank test for related samples. The differences between paired observations are not normally distributed, so this method is preferred over the $t$-test.}
\label{tab:testDiff}
\end{table}

As an additional exercise, we illustrate how the set of complementary policy issues with spillover effects is country-specific. For this, consider the 13 development pillars, and their corresponding policy issues. Taking only the negative and significant $\beta$ coefficients that have links to the RoL, we construct the total $\beta$ coefficient of development pillar $j$ as $\sum_c \sum_{i \in P_{j}} |\beta_{i,c}|$, where the outer sum runs across countries in a specific cluster and the inner across policy issues belonging to pillar $j$. The reason to take a sum instead of an average is to account for the fact that some policy issues are complementary to the RoL in some countries but not in others, so if an issue appears more frequently across countries, it becomes more important. Hence, this exercise is mainly illustrative and has the purpose of facilitating a comparison between development pillars.

\begin{figure}[ht]
    \centering
    \caption{Total $\beta$ coefficients for policy issues with links to the RoL}
    \includegraphics[scale=.55]{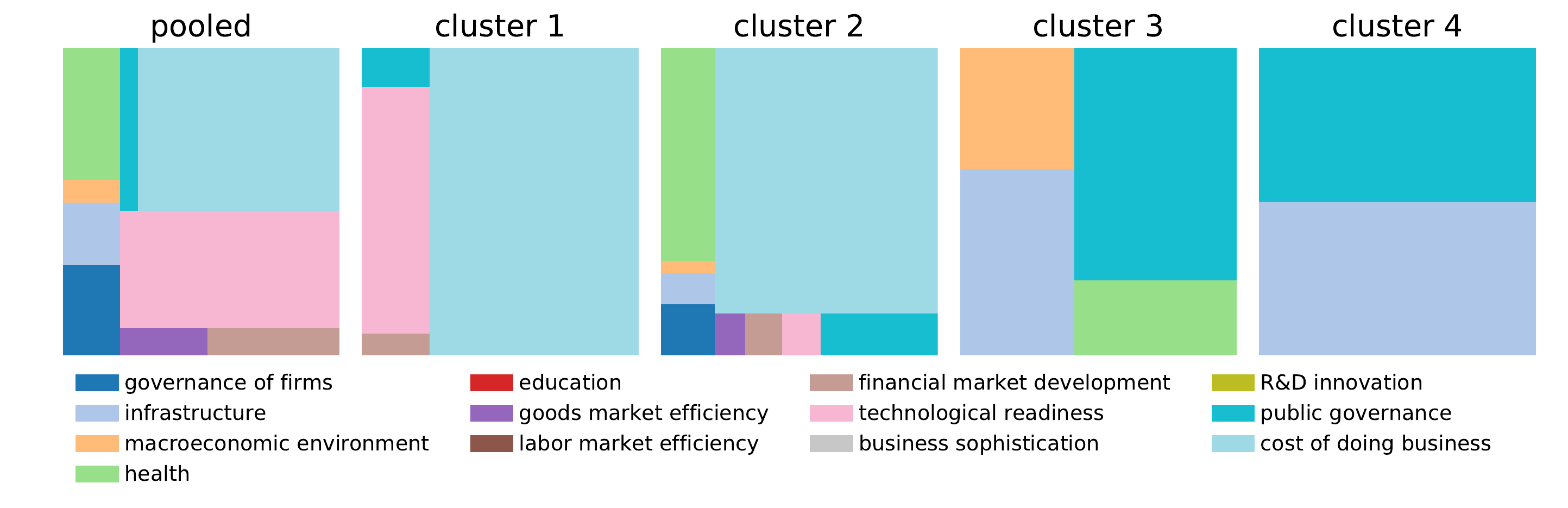}
    \label{fig:treemap_link}
\end{figure}

Figure \ref{fig:treemap_link} shows treemap plots for the pooled dataset and for each of the clusters.\footnote{When performing the same procedure to negative-significant beta coefficients without links to the RoL, the composition of complementary policy issues looks rather homogeneous across clusters (see appendix \ref{app:complementarities}).} In general, the composition of complementary topics is strikingly different. For instance, improvements to \emph{public governance} are considerably more complementary to the RoL in less developed countries. In contrast, reducing the \emph{cost of doing business} seem to be more useful in high and middle-high income countries (clusters 1 and 2). This confirms the importance of producing country-specific estimates in order to build bespoke policy profiles.

\subsection{Effective policy profiles}

Our last set of results explores whether policy profiles that successfully reduce corruption tend to prioritize the RoL relative to other policy issues. We can think of such prioritization in two ways: (1) priority in terms of goals through $T$ and (2) priority in terms of allocated resources through $P$. For this we test hypotheses IV.a and IV.b.

First, we generate target vectors $T'$ under which $\bar{D}$ is lower than the empirically estimated under the retrospective analysis.\footnote{Since the space of possible target vectors is too large, we employ a heuristic optimization algorithm called \emph{differential evolution} \citep{storn_differential_1997}. For a given country, the algorithm runs until it finds a $T'$ such that $\bar{D}$ drops by 25\% with respect to the empirical level. Then, for that country, we recover all those vectors that also produced reductions in $\bar{D}$ during the optimization process.}. Second, in order to evaluate relative priorities, we calculate the rank $Q_T$ that the RoL has in the empirical target vector $T$ (1 being the highest entry in $T$ and 79 the lowest). Then, the difference $Q_T - Q_T'$ represents the relative change in the priority that the RoL receives among the targets (a positive sign means an improvement in the ranking). Alternatively, $Q_P - Q_P'$ denotes the relative change in priority in terms of the allocated resources. We employ these two metrics to test hypotheses IV.a and IV.b respectively. A positive correlation between $\bar{D}-\bar{D}'$ and $Q_T - Q_T'$ would suggest that, among the policy profiles that effectively dampen corruption, larger reductions are associated to higher relative priorities to the RoL target, supporting hypothesis IV.a. This, however, is not the case. We find that the Spearman correlation between these metrics is -0.01 with a $p$-value of 0.73. Similarly, it is -0.062 with a $p$-value of 0.04 for allocative priorities. Therefore, our results do not support any of the hypotheses from family V.

\begin{figure}[ht]
    \centering
    \caption{Relative changes in the priority of the RoL}
    \includegraphics[scale=.75]{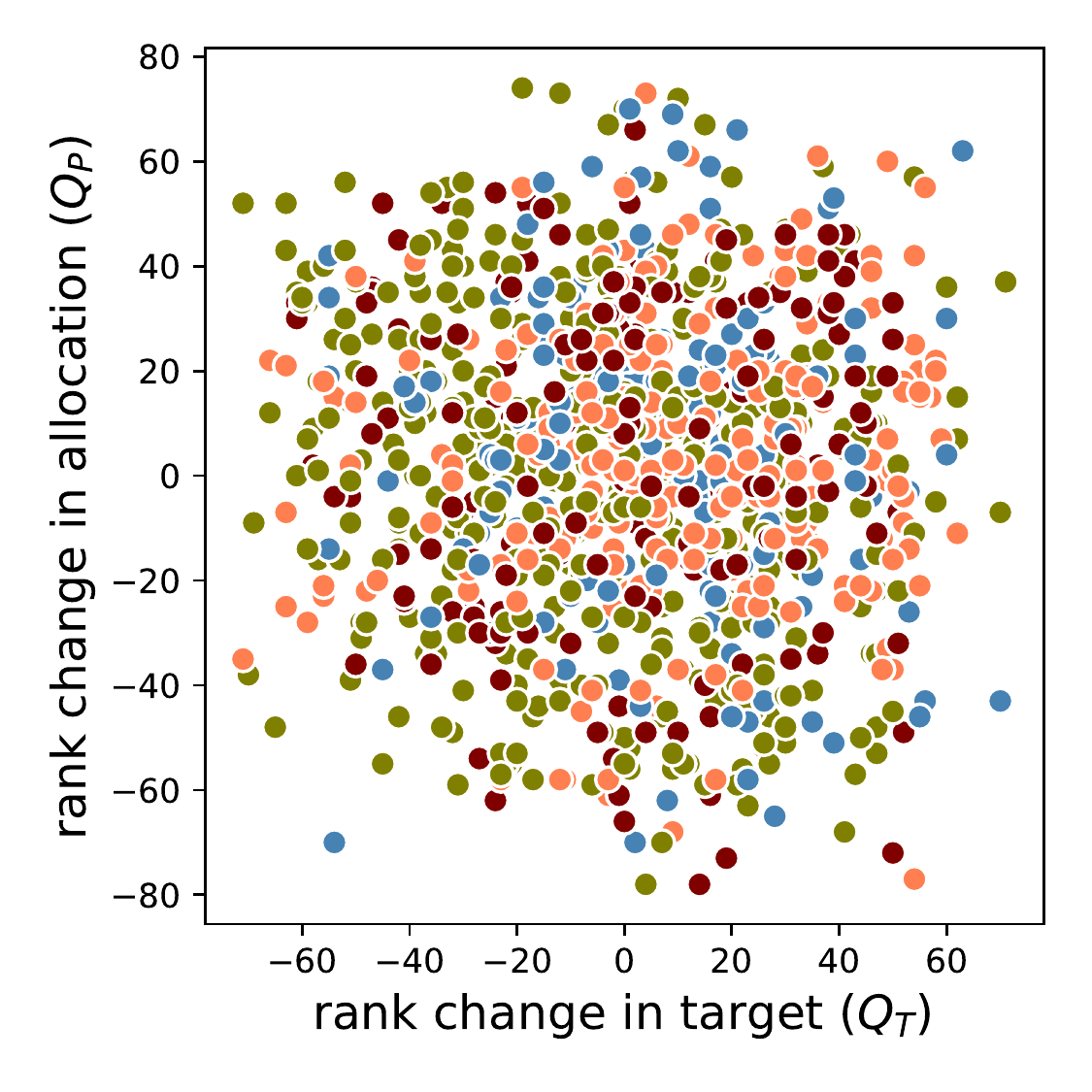}
    \caption*{The color of the dots corresponds to the cluster identification: blue for 1, green for 2, orange for 3 and maroon for 4.}
    \label{fig:ranks}
\end{figure}

Figure \ref{fig:ranks} summarizes these results by presenting a scatter plot combining target and allocative changes in priorities. The main take-away is that effective anti-corruption prescriptions that encourage improvements to the RoL do not imply a higher prioritization of this topic over other policy issues. In fact, a complete set of priorities can only be derived through a systemic view of the problem. Thus, overemphasizing the RoL and turning it into the most important agenda could become a misleading endeavour.\footnote{Paradoxically, a country positioned in the negative quadrant can become more efficient with the proper policy profile, even if its RoL priorities (in targets and allocations) are reduced with respect to the empirical ones.}

\section{Conclusions}\label{sec:conclusions}

Our results suggest that a rule-of-law heavily-oriented governance agenda, by itself, will hardly achieve the desired low-corruption outcomes. In particular, isolated reforms to the rule of law (inspired solely on a principal-agent view) that reshape incentives and restrain opportunities for discretionary expenses are likely to be ineffective, especially in the developing countries. In our view, this is a consequence of neglecting the systemic features of corruption such as spillovers between public policies and social norms that evolve in an uncertain environment. Therefore, in order for policy profiles to be truly effective, it is important to embed some type of coordinating device that can guide the system towards low-corruption norms. In such profiles, special importance should be given to policy issues that lie beyond the governance agenda and that could be complementary in the abatement of corruption.

Our computational approach allows producing country-specific estimates on the effectiveness of the RoL under different settings in relation to other covariates.  Simulation outcomes support the importance of country-specific context and, thus, advocate for policy prescriptions that are tailored to satisfy the particular and distinct needs of each government. In such prescriptions, improvements to the RoL are necessary but not substitutes of improvements in other policy issues. That is, prescriptions that emphasize improvements to the RoL should be careful of not over-prioritizing this topic (at least not if the objective is reducing corruption). Clearly, the complexity of this problem and the instruments to tackle it require alternative methods than can deal with systemic attributes. Such tools can enable the discovery of counter-intuitive explanations and the production of ex-ante evaluations to analyze the effectiveness of alternative policy tools.

\bibliography{references}

\appendix

\section{Countries and clusters}\label{app:countries}

Table \ref{tab:list_countries} provides a list of all the countries that are included  in each cluster. Countries are identified through their International Organization for Standarization code (alpha-3).

\begin{table}[!htb]
   \linespread{0.95} \centering\footnotesize
    \caption{List of countries by cluster}\label{tab:list_countries}
    \begin{tabular}{p{3cm}cp{8cm}}
        \hline
        \textbf{Cluster} & \textbf{Number of countries} & \textbf{Countries}\\\hline
        1) High  & 24 &
        ARE AUS AUT BEL CAN CHE DEU DNK FIN FRA GBR HKG IRL ISR JPN KOR MYS NLD NOR NZL QAT SGP SWE USA \\\hline
        2) Mid-high  & 45 &
        BHR BRA BWA CHL CHN COL CRI CZE EGY ESP EST GRC GTM HND HRV HUN IDN IND ITA JAM JOR KWT LKA LTU LVA MAR MEX MUS NAM OMN PAN PER PHL POL PRT SAU SLV SVK SVN THA TTO TUN TUR URY ZAF \\\hline
        3) Mid-low  & 24 &
        ALB ARG ARM AZE BGD BGR BIH BOL DOM DZA ECU GEO KAZ KGZ MKD MNG NIC NPL PRY RUS SRB TJK UKR VEN \\\hline
        4) Low  & 24 &
        BDI BEN BFA CIV CMR ETH GHA GMB KEN KHM MDG MLI MOZ MRT MWI NGA PAK SEN TCD TZA UGA VNM ZMB ZWE \\\hline
    \end{tabular}
\end{table}

\section{Co-evolutionary learning}\label{app:equations}

In this appendix, we present the equations modeling the public functionaries' learning process. This variant of reinforcement learning helps them to determine her/his levels of contribution to the investment in policy issue $i$  

\begin{equation}
    C_{i,t} = \min \left\{ P_{i,t},
     \max\left(0, C_{i,t-1} +  d_{i,t} |\Delta F_{i,t}| \frac{C_{i,t-1} + C_{i,t-2}}{2} \right) \right\} \label{eq:contribution},
\end{equation}
where $\Delta F_{i,t}$ is the most recent change in benefits and $d_{i,t}$ is the sign function

\begin{equation}
    d_{i,t} = \text{sgn} (\Delta F_{i,t} \cdot \Delta C_{i,t}),\label{eq:sign}
\end{equation}
such that

\begin{equation}
    \begin{split}
        \Delta F_{i,t} = F_{i,t-1} - F_{i,t-2}\\
        \Delta C_{i,t} = C_{i,t-1} - C_{i,t-2}.
    \end{split}
\end{equation}

That is, through direct signals obtained from their gains and losses of benefits with respect to the previous period, these bureaucrats learn to maintain or reverse a trend in contribution changes. In other words, they keep the direction of the change when observing an increase in benefits, and modify it when observing a reduction.

\section{Treemap plots for non-complementary policy issues}\label{app:complementarities}

\begin{figure}[ht]
     \centering
     \caption{Total $\beta$ coefficients for policy issues without links to the RoL}
     \includegraphics[scale=.55]{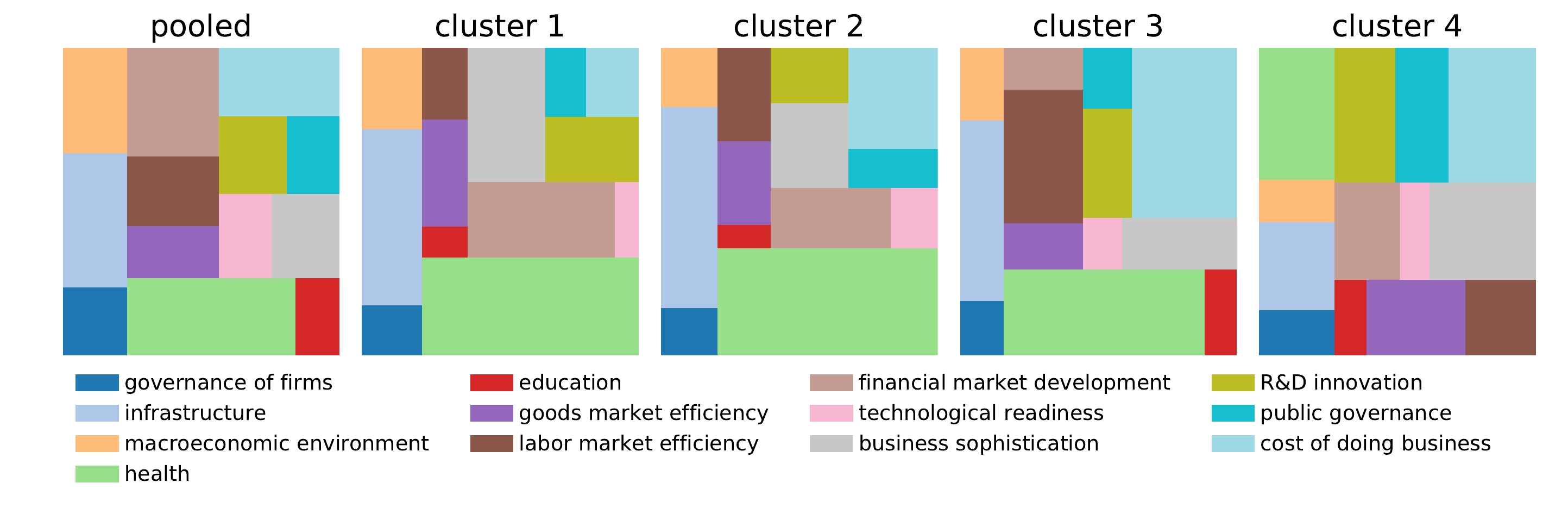}
     \label{fig:treemap_noLink}
 \end{figure}

Note that, in Figure \ref{fig:treemap_noLink}, a wide diversity of policy issues with no links to RoL exhibit a negative impact on corruption. However, there is not a notorious heterogeneity in the magnitude of these impacts across clusters.    

\end{document}